\documentclass[11pt,reqno]{amsart}
\usepackage[hidelinks]{hyperref}
\usepackage{packages, mathtools}

\newcommand{\bW}{\mathbb{W}}

\newcommand{\hol}{\mr{hol}}
\renewcommand{\s}{\sigma}

\newcommand{\I}{\mc{I}}
\newcommand{\Dx}{\Delta x}
\theoremstyle{plain}
\renewcommand{\thefigure}{\Roman{figure}}

\newtheorem{innercustomthm}{Theorem}
\newenvironment{customthm}[1]
  {\renewcommand\theinnercustomthm{#1}\innercustomthm}
  {\endinnercustomthm}

\begin{document}

  \title[Stochastic Feynman Rules]{Stochastic Feynman Rules for\\Yang-Mills Theory on the Plane}
  \author{Timothy Nguyen}
  \email{timothy.c.nguyen@gmail.com}
\date{\today}

\begin{abstract}
We analyze quantum Yang-Mills theory on $\R^2$
using a novel discretization method based on an algebraic analogue of stochastic calculus. Such an analogue involves working with ``Gaussian'' free fields whose covariance matrix is indefinite rather than positive definite. Specifically, we work with Lie-algebra valued fields on a lattice and exploit an approximate gauge-invariance that is restored when taking the continuum limit. This analysis is applied to show the equivalence between Wilson loop expectations computed using partial axial-gauge, complete axial-gauge, and the heat-kernel lattice formulation.  As a consequence, we obtain intriguing Lie-theoretic identities involving heat kernels and iterated integrals.
\end{abstract}

\maketitle

\tableofcontents

\section{Introduction}

Quantum Yang-Mills theory on $\R^2$ is exactly soluble due to the fact that the theory becomes free in (complete) axial-gauge. More precisely, working in a gauge in which the connection $A = A_xdx + A_ydy$ has $A_y \equiv 0$ and $A_x$ vanishes on the $x$-axis, the Yang-Mills action becomes purely quadratic. This allows for a rigorous interpretation of the Yang-Mills measure as a Gaussian measure, or more precisely, a (Lie-algebra valued) white-noise measure \cite{Dri}. As a result, Wilson loop expectation values can be analyzed in terms of stochastic holonomy \cite{GKS}, whereby random connections are distributed according to the Yang-Mills measure. These resulting stochastic computations of \cite{Dri} agree with formulas arising from the heat-kernel lattice formulation\footnote{There is also the Wilson lattice formulation in arbitrary dimension. In dimension two, its continuum limit recovers the heat kernel lattice formulation, see \cite{ALMMT}.} of Yang-Mills theory \cite{Mig} which uses the heat kernel action and which has been extensively studied from a variety of perspectives \cite{ALMMT, CMR, DGHK, Fine1, Levy, Sen, Sin}. In contrast to \cite{Dri} however, the standard method of evaluating expectation values of observables against Gaussian measures is to apply Wick's Theorem, i.e., to evaluate and sum over Feynman diagrams. The Feynman diagrammatic method has the advantage that it readily carries over to interacting quantum field theories, whereas the exact (i.e. nonperturbative) methods using white-noise analysis does not. Moreover, the conceptual requirements needed to implement Feynman diagrams, such as those pertaining to gauge-fixing and gauge-invariance, provide a rich and important arena for mathematical analysis. Among the gauge-fixing procedures most relevant to two-dimensional Yang-Mills theory are the complete axial-gauge and the partial axial-gauge, where the latter imposes that only $A_y$ vanish. (Partial axial-gauge has the advantage that translation-invariance is manifestly preserved, whereas it must be established with considerable work in the case of complete axial-gauge \cite{Dri, GKS}.)


The purpose of this paper is to show that Wilson loop expectations computed using Feynman diagrams in complete and partial axial-gauge agree with those computed using stochastic/lattice methods. In particular, partial axial-gauge and complete axial-gauge are equivalent. From a physical standpoint, this is to be expected since physical quantities should be independent of the choice of gauge. Mathematically however, such reasoning cannot be applied so straightforwardly owing to the fact that the use of gauge-transformations when working with low regularity random connections requires considerable care. In fact, the main difficulty  runs much deeper, since the partial axial-gauge does not even give rise to an honest Gaussian measure since the connection is massless. Our initial attempt to circumvent this problem was to insert a mass regulator and then send the mass to zero, but this turns out to introduce difficulties we were not able to surmount, see Remarks \ref{Rem:Massless} and \ref{Rem:c}. So instead, we forgo measure theory and directly consider computations in the partial axial-gauge as defined purely algebraically using the Wick procedure. In this way, we are naturally led to develop an algebraic analogue of (discretized) white-noise analysis and stochastic calculus that is able to handle the partial axial-gauge. The use of gauge-invariance in this setting turns out to require significant finesse, as will become apparent later on. Our final result, which computes Wilson loop expectations in two different ways, is in purely mathematical terms a collection of nontrivial identities between integrals of heat kernels on the gauge group with iterated integrals along contours defining our Wilson loops. The ability to evaluate iterated integrals plays an important role in a variety of mathematical and quantum field theoretic settings \cite{BrownII}. Our own motivation stems from an investigation into the fundamental mathematical structure of quantum gauge theories, see the discussion at the end of the introduction.

We now describe our main results, with a more complete discussion of the setup given in Section \ref{Sec:2}. Fix any compact Lie group $G$ for our gauge group. A Wilson loop observable $W_{f,\gm}$ takes a (sufficiently smooth) connection $A$, computes its holonomy $\hol_\gm(A)$ about a (piecewise $C^1$) closed curve $\gm$, and then applies the conjugation-invariant function $f: G \to \bC$ to this group-valued element:
\begin{equation}
 W_{f,\gm}(A) := f(\hol_\gm(A)).
\end{equation}
Using the path-ordered exponential representation of $\hol_\gm(A)$, we can express $W_{f,\gm}(A)$ as a power series functional in $A$ via (\ref{eq:WLexp}).

From this representation, we can compute Wilson loop expectations, term by term, with respect to the (putative) Yang-Mills measure in partial or complete axial-gauge. This computation makes use of the Wick rule, which determines the expectation of any polynomial from the two point function (i.e. the expectation of a quadratic polynomial), see (\ref{eq:Wick_rule}). The use of the Wick rule means that we do not need an honest measure to compute expectations, though the expectation comes from a Gaussian measure in case the two-point function defines a positive-definite pairing. The two-point function in partial and complete axial-gauge is obtained as follows. First, we determine the corresponding (gauge-dependent) Green's functions for the Yang-Mills kinetic operator $-\pd_y^2$. Next, the integral kernels of these Green's functions give rise to corresponding propagators $P_{pax}$ and $P_{ax}$ in partial and complete axial-gauge, respectively, see (\ref{eq:Ppax}--\ref{eq:Pax}).
Finally, the insertion of these propagators into bilinear expressions of the connection $A$ determines the two-point function, i.e., the ``Feynman rules''.

From these propagators, we can attempt to make the following formal definitions\\
\begin{align}
\begin{split}
 \left<W_{f,\gm}\right>_{pax},\; \left<W_{f,\gm}\right>_{ax} ``=\textrm{''}\; &\textrm{sum over all Feynman integrals obtained from}\\
 &\textrm{inserting propagators } P_{pax} \;\textrm{(resp. } P_{ax}) \textrm{ weighted }  \\
 &\textrm{by $\lambda$ into (\ref{eq:WLexp}) using the Wick rule.}\end{split} \label{eq:WLE}
\end{align}
(For those unfamiliar with the Feynman diagram  procedure implicit on the right-hand side above, see  (\ref{eq:WLEformula}) for an explicit formula.) The weight $\lambda$ is a parameter which counts the number of propagator insertions; in path integral notation, it coincides with the Yang-Mills coupling constant in (\ref{eq:YMpax}).

The reason the above ``definition'' is formal is that the propagators (\ref{eq:Ppax}) and (\ref{eq:Pax}) are singular and so do not a priori yield well-defined integral expressions in the evaluation of (\ref{eq:WLE}). This turns out not to be a problem for partial axial-gauge, since the $y$-dependent part of $P_{pax}$ vanishes along the diagonal and so ``cancels'' a $\delta$-function in the $x$-direction. But this being not the case for $P_{ax}$, the definition (\ref{eq:WLE}) is ill-defined because while the classical integrals occuring in (\ref{eq:WLexp}) are given by Riemann integrals of a smooth connection that do not depend on how the integral is discretized, the quantum expectation values of these integrals in complete axial-gauge \textit{depends on the discretization method}. This is  most easily illustrated by noting that
\begin{equation}
\int_{1 > t_2 > t_1 > 0}\delta(t_2-t_1)dt_2dt_1 = 0 \label{eq:delta1}
\end{equation}
whereas
\begin{equation}
\int_{1 \geq t_2 \geq t_1 \geq 0}\delta(t_2-t_1)dt_2dt_1 = \int_0^1 dt_1 = 1. \label{eq:delta2}
\end{equation}
Hence, owing to the fact that the integrand $\delta(t_2-t_1)$ is singular, how one approximates the domain of integration $\{1 \geq t_2 \geq t_1 \geq 0\}$, whether from below or above or an average of the two, affects the result of the integration. (Though if one weights the above integrands with a continuous function that vanishs at $t_1=t_2$, there would be no ambiguity.)

The resolution of this predicament is that the Wilson loop expectation in complete axial-gauge is not an operation on an underlying classical observable (in this case a Riemann integral). Rather, the Wilson loop expectation is an expectation of a \textit{stochastic integral}. There are two common constructions for stochastic integrals: the It\^{o} integral, which uses the left-endpoint rule and the Stratonovich rule which uses the midpoint rule. They yield different answers in a manner analogous to the above computation.

Consequently, the most natural way to define $\left<W_{f,\gm}\right>_{ax}$ is by regarding the integrals occurring in the path-ordered exponential (\ref{eq:WLexp}) as iterated Stratonovich integrals of the nonsmooth random connections $A$. These integrals then become random matrix-valued elements, from which we can apply $f = \tr$ and take the (stochastic) expectation to obtain the final numerical result. The use of the Stratonovich integral is most natural since this integral obeys the usual calculus rules under changes of variables and the like (the It\^{o} integral does not).

We now have three ways to compute Wilson loop expectations. We have $\left<W_{f,\gm}\right>_{pax}$ and $\left<W_{f,\gm}\right>_{ax}$, both of which are computed ``Feynman diagramatically''. (For the complete axial-gauge, the expectation of iterated Stratonovich integrals can be computed by a Wick procedure which converts Stratonovich integrals to It\^{o}-Riemann integrals, see Lemma \ref{Lemma:StoI}.) They are a priori formal power series in the coupling constant $\lambda$. We also have the exact expectation $\left<W_{f,\gm}\right>$, which is computed using the heat kernel action on a lattice. It is the exact expectation because such a lattice formulation is invariant under subdivision and so represents an exact expectation of $W_{f,\gm}$ in the continuum limit. The quantity $\left<W_{f,\gm}\right>$ is a well-defined function of $\lambda$, for $\lambda > 0$. For us, we can take the stochastic expression (\ref{eq:Dri}) as the definition of $\left<W_{f,\gm}\right>$ (the work of \cite{Dri} implies that this definition is equivalent to the more commonly used definition in terms of heat kernels).

Our first result amounts to a basic unraveling of stochastic constructions:

\begin{customthm}{1}\label{Thm:1}
 We have $\left<W_{f,\gm}\right>_{ax}$ defines an entire\footnote{
Here, we assume that $f$ is trace in an irreducible represention pf $G$ (or more generally is a polynomial in such functions) to assert that $\left<W_{f,\gm}\right>_{ax}$ is entire. For a general conjugation-invariant function $f$, one needs some kind of restriction on $f$ since an infinite linear combination of entire power series may not result in one with a positive radius of convergence.} power series in $\lambda$ and 
 \begin{equation}
  \left<W_{f,\gm}\right> = \left<W_{f,\gm}\right>_{ax}
 \end{equation}
  in the sense that the two functions agree for $\lambda > 0$.
\end{customthm}

Our main result however concerns the equivalence between partial axial-gauge and complete axial-gauge:

\begin{customthm}{2}\label{Thm:2}
 We have
 \begin{equation}
  \left<W_{f,\gm}\right>_{pax} = \left<W_{f,\gm}\right>_{ax}. \label{eq:maineq}
 \end{equation}
\end{customthm}
\noindent The above results and their proofs go through unchanged if we consider products of Wilson loop observables.

Our proof of Theorem 2 requires new ideas. This is because as mentioned before, partial axial-gauge does not arise from a measure since the corresponding two-point function defines an indefinite pairing. The approach we develop is to simply push stochastic analytic constructions in the indefinite setting. This requires a pscyhological adjustment, akin to working with anticommuting Grassman variables instead of ordinary commuting variables. In the indefinite setting, ``random variables'' no longer have values, i.e. they are not functions defined on a measure space. Instead, they form an algebra that is equipped with an expectation operator defined by use of the Wick rule. Hence, we regard the analysis we develop as an \textit{algebraic stochastic calculus}. A notable feature of this calculus is that because the associated two-point function (i.e. covariance) is no longer positive, one must forgo most basic tools for estimates such as the Cauchy-Schwarz inequality. While there exist other extensions of classical probability theory \cite{BS, HP}, it is unclear what (if any) relations these constructions might bear with ours.\\

\noindent\textbf{Example: } We provide the simplest explicit example of the equality $\left<W_{f,\gm}\right> = \left<W_{f,\gm}\right>_{pax}$ in order to illustrate the type of identities our theorems provide. These identities become highly nontrivial as the complexity of $\gm$ increases.

Let $f = \tr \rho(\cdot)$ be trace in some irreducible representation $\rho$ of our gauge group $G$. Let $\gm$ be a curve given by joining two disjoint horizontal curves (see Definition \ref{Def:horizontal}) by vertical segments, see Figure 1. Let $R$ denote the region it encloses and $|R|$ its area. In this case, we can easily verify (\ref{eq:maineq}) to all orders in $\lambda$ as follows:

\renewcommand{\thefigure}{\arabic{figure}}

\begin{figure}\label{Fig1}
 \vspace{.2in}\includegraphics[scale=0.8]{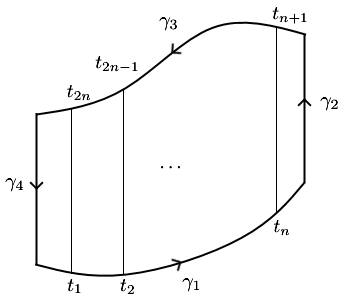}
 \caption{Propagators joining $(t_1, t_{2n}), (t_2, t_{2n-1}), \ldots, (t_n,t_{n+1})$ in the order $\lambda^n$ term of $\left<W_{f,\gm}\right>_{pax}$.}
\end{figure}

To compute $\left<W_{f,\gm}\right>_{pax}$,  write $\gm: [0,1] \to \R^2$ as the concatenation of paths $\gm_4.\gm_3.\gm_2.\gm_1$, where (see Figure \ref{Fig1})
\begin{itemize}
 \item $\gm_1$ is the graph of $\gm_-: [t_0,T] \to \R$ traversed from $t_0$ to $T$;
 \item $\gm_3$ is the graph of $\gm_+: [t_0,T] \to \R$ traversed from $T$ to $t_0$;
 \item $\gm_-(t) < \gm_+(t)$ for all $t$;
 \item $\gm_2$ and $\gm_4$ are upward and downward moving vertical segments, joining the $\gm_1$ and $\gm_3$.
\end{itemize}
In computing $\left<W_{f,\gm}\right>_{pax}$, the order $\lambda^n$ term of $\left<W_{f,\gm}\right>_{pax}$ comes from inserting $n$ copies of $P_{pax}$ given by (\ref{eq:Ppax}) into the order $2n$ term of (\ref{eq:WLexp}). When we Wick contract a pair of points on $\gm$ at times $t_i < t_j$, we only pick up a nonvanishing term when we contract a point of $\gm_1$ with that of $\gm_3$. (In partial axial-gauge, the propagator has only has $dx$-components and so integrates trivially along vertical segments; moreover, the $\delta$-constraint in the $x$-direction means we cannot contract pairs of points belonging to the same $\gm_i$). Each such Wick contraction yields both the term
\begin{equation}
-\left[-\frac{|\gm_+(x(t_j)) - \gm_-(x(t_i))|}{2}\delta(x(t_i)-x(t_j))\right], \label{eq:prop-ex}
\end{equation}
where $x(t)$ denotes the $x$-coordinate of $\gm(t)$, and an insertion of $e_ae_a$ (corresponding to the identity tensor on $\g$, where $e_a$ is an orthonormal basis for the Lie algebra $\g$ and we implicitly sum over the repeated index $a$). We pick up a factor of $-1$ in (\ref{eq:prop-ex}) since $\gm_3$ and $\gm_1$ go in opposite directions.

As a result, we need to choose all possible groupings of the $2n$ path-ordered points $t_1,\ldots, t_{2n}$ into $n$ pairs, each such pair determining the location of a propagator insertion. The $\delta$-constraint in the $x$-direction in the propagator means that the only nonvanishing pairings arise from the choice $(t_1, t_{2n}), (t_2, t_{2n-1}), \ldots, (t_n, t_{n+1})$, with $t_1,\ldots, t_n$ belonging to the domain of $\gm_1$ and $t_{n+1}, \ldots, t_{2n}$ belonging to the domain of $\gm_3$. Thus the $\lambda^n$ coefficient of $\left<W_{f,\gm}\right>_{pax}$ is given by
\begin{multline*}
\begin{split}
\frac{1}{2^n}\int_{1 \geq t_{2n} \geq t_{2n-1} \geq \cdots \geq t_1 \geq 0}\bigg[\Big(\gm_+(x(t_{2n})) -\gm_-(x(t_1))\Big)\cdots \Big(\gm_+(x(t_{n+1})) -\gm_-(x(t_n))\Big)\bigg] \times \\
\bigg[\delta(x(t_{2n})-x(t_1))\cdots \delta(x(t_{n+1})-x(t_n))\bigg]\tr\Big(\rho(e_{a_1})\cdots \rho(e_{a_n})\rho(e_{a_n})\cdots \rho(e_{a_1})\Big)\prod dt_i
\end{split}\\
\begin{split}
& = \frac{[\tr(\rho(e_a)\rho(e_a))]^n}{2^n}\int_{T \geq x_n \geq \cdots \geq x_1 \geq t_0}(\gm_+(x_1) - \gm_-(x_1))\cdots (\gm_+(x_n)-\gm_-(x_n)) \prod dx_i \\
& = \frac{c_2(\rho)^n}{2^n}\frac{1}{n!}\int_{t_0}^T (\gm_+(x)-\gm_-(x))^ndx \\
& = \frac{c_2(\rho)^n|R|^n}{2^nn!},
\end{split}
\end{multline*}
where $c_2(\rho) = \tr(\rho(e_a)\rho(e_a))$ denotes the quadratic Casimir for the representation $\rho$. In going from the first line to the second, we used that $e_ae_a$ is a central element in the universal enveloping algebra of $\g$.

On the other hand, we have
\begin{align}
 \left<W_{f,\gm}\right> &= \int_G \tr\, \rho(g) K_{\lambda|R|}(g)dg \label{eq:WLEexample}\\
 &= e^{-\lambda|R| c_2(\rho)/2}
\end{align}
where $K_t(g)$ is the convolution kernel for the heat operator $e^{-t\Delta/2}$ on $G$ (with respect to Riemannian metric induced by the inner product on $\g$). Indeed, the rule for computing $\left<W_{f,\gm}\right>$ involves placing a heat kernel at every bounded face of $\R^2\setminus\gm$ at time equal to the coupling constant $\lambda$ multiplied by the area enclosed \cite{Dri, Levy}. The argument of each heat kernel and of $f$ is formed from words formed out of group elements labeling the edges of $\gm$. In the simplest case of a simple closed curve, we can treat all of $\gm$ as a single edge and so we have just a single group element $g \in G$ in (\ref{eq:WLEexample}), and $f$ and $K_{\lambda|R|}$ are evaluated on $g$.

Thus, for every $n$, the order $\lambda^n$ terms of $\left<W_{f,\gm}\right>$ and $\left<W_{f,\gm}\right>_{pax}$ agree. The explicit evaluation of $\left<W_{f,\gm}\right>_{ax}$ is a bit more involved. See Remark \ref{Rem:ax} for details.\\

\noindent\textit{Outline of paper and further remarks:}\\

This paper is organized as follows. In Section 2, we describe a more detailed setup of the axial gauges and their propagators, as well as some terminology concerning our decomposition of Wilson loops into simpler pieces. We also discuss how maps generalizing white-noise naturally arise when performing parallel transport in axial gauges. Since this generalization involves forgoing the usual positive-definiteness conditions in probability theory, in Section 3 we develop an algebraic stochastic calculus to analyze such generalized white-noise. It involves a discretization procedure which approximates stochastic integrals by finite sums, with the latter being well-defined when forgoing measure theory. We should emphasize that our goal here is not a systematic development of this calculus, but rather to present ideas that are best motivated by such introducing such a calculus. Finally, we apply these results to prove our main theorems in Section 4.

We would like to make some remarks on how our work fits into the greater context of quantum field theory. First, we note that our discretization method appears to be new. The standard approach to discretizing gauge theories involves placing group valued, not Lie-algebra valued fields, on a lattice. Such a discretization has a built in gauge-invariance. For us, our use of gauge-invariance only holds asymptotically, i.e., it is restored in the continuum limit. Next, we make a general observation concerning the dichotomy between two different approaches to quantum field theory. On the one hand, there has been the long-standing constructivist school which aims to construct honest measures for defining path integrals. On the other hand, because this task is plagued with many difficulties, it has been fashionable for many decades to instead understand what kind of mathematics one can derive from purely formal aspects of path integrals (with the many works of Witten being the pinnacle of such endeavors). A notable aspect of our work is that it provides a kind of mysterious link between the formal (not arising from a measure) aspects of partial axial-gauge to the measure-theoretic aspects of complete axial-gauge. We believe this connection deserves to be better understood. Finally, we believe the algebraic stochastic calculus we formulated is worthy of being further developed, not only for its own mathematical sake, but because it may be useful in other quantum field theoretic settings in which one has two-point functions that are not positive.

This paper is an output of the author's investigation into the relationship between the perturbative and the exact formulation of two-dimensional Yang-Mills theory \cite{Ngu-YM, Ngu2016}. In these works, holomorphic gauge instead of partial axial-gauge is studied, where holomorphic gauge is regarded as a ``generalized axial-gauge'' distinct from the axial gauges considered here\footnote{One can regard the partial axial-gauge and complete axial-gauges here as ``stochastic axial-gauge'', since the two gauges are equivalent and complete axial-gauge involves a ``stochastic regulator''. On the other hand, generalized axial-gauge is a generalization of Wu-Mandelstam-Liebrandt light cone gauge, which regulates partial axial-gauge using a different regulator. See \cite{Ngu-YM, Ngu2016} for further details.}. In fact,
it is shown that holomorphic gauge is inequivalent to the axial gauges considered here, which suggests that our main theorems concerning various equivalences are not to be taken for granted. Furthermore, \cite{Ngu2016} shows that Wilson loop expectations in holomorphic gauge also yield a set of remarkable identities relating iterated integrals and matrix integrals. We hope the stochastic analysis we provide here, which yields identities to all orders in perturbation theory, may be useful in other (quantum field-theoretic) contexts.

\section{The Setup}\label{Sec:2}

Fix a compact Lie group $G$ equipped with an ad-invariant inner product $\left<\cdot,\cdot\right>$ on its Lie algebra $\g$. On $\R^2$, given a (smooth) connection $A = A_xdx + A_ydy$, which is an element of $\Omega^1(\R^2;\g)$, the space of $\g$-valued $1$-forms, we can always find a gauge transformation that places $A$ in \textit{partial axial-gauge}:
\begin{equation}
 A \in \mc{A}_{pax} = \{A \in \Omega^1(\R^2;\g) : A_y \equiv 0\}.
\end{equation}
Having done so, we still have gauge freedom in the $x$-direction to place a connection in \textit{complete axial-gauge}:
\begin{equation}
A \in \mc{A}_{ax} = \{A \in \Omega^1(\R^2;\g) : A_y \equiv 0,\; A_x(\cdot,0) \equiv 0\}.
\end{equation}
This completely eliminates all gauge freedom arising from the action of the group of gauge transformations that are fixed to be the identity at the origin.

In either of these gauges, the (Euclidean) Yang-Mills path integral can be formally written as\footnote{The Faddeev-Popov determinant is constant in axial gauges and hence can be dropped in the path integral.}
\begin{equation}
\int dA_x e^{-\frac{1}{2\lambda}\int dxdy\,\left<\pd_yA_x, \pd_yA_x\right>}. \label{eq:YMpax}
\end{equation}
where $A_x$ ranges over either $\A_{pax}$ or $\A_{ax}$ and $\lambda$ is a coupling constant. Indeed, in the these axial gauges, the curvature of the connection $A$ is simply $\pd_y A_x dy\wedge dx.$
The integrand of (\ref{eq:YMpax}) is the putative Yang-Mills measure.

The Wick rule we obtain from (\ref{eq:YMpax}) is determined by the Green's function we choose for the kinetic operator $-\pd_y^2$ occuring in (\ref{eq:YMpax}). A Green's function satisfies
\begin{equation}
 -\pd_y^2G(x,y; x',y') = \delta(x-x')\delta(y-y'). \label{eq:Green}
\end{equation}
For partial axial-gauge, we choose the unique solution to (\ref{eq:Green}) which is invariant under translations and reflections and homogeneous under scaling. Indeed, these are symmetries of the kinetic operator, and so we may as well impose them on the Green's function. We obtain
\begin{align}
 G_{pax} =\bar G_{pax}(y,y')\delta(x-x')
\end{align}
where
\begin{align}
 \bar G_{pax}(y,y') &= -\frac{|y-y'|}{2}.
\end{align}
For complete axial-gauge, we consider $-\pd_y^2$ acting on functions which vanish along the axis $y=0$. We impose the same symmetries as before, only now we must relinquish translation invariance in the $y$-direction. We thus obtain the Green's operator
\begin{equation}
\bar G_{ax}(y,y') = \begin{cases}
             \min(|y|,|y'|) & yy' \geq 0 \\
             0 & yy' < 0.
            \end{cases} \label{eq:tG}
\end{equation}
From the partial axial-gauge and complete axial-gauge Green's function, we obtain a corresponding \textit{propagator}, which promotes these scalar Green's functions to elements of $\A_{pax} \otimes_\R \A_{pax}$ and $\A_{ax} \otimes_\R \A_{ax}$ respectively:
\begin{align}
 P_{pax}(x-x',y-y') &= -\frac{|y-y'|}{2}\delta(x-x')(dx \otimes dx')  e_a\otimes e_a \label{eq:Ppax} \\
 P_{ax}(x-x';y,y') &= \bar G_{ax}(y,y')\delta(x-x')(dx \otimes dx') e_a\otimes e_a \label{eq:Pax}
\end{align}
These propagators are integral kernels of Green's operators (with respect to the inner product pairing on $\g$-valued forms) for $-\pd_y^2$ acting on $\A_{pax}$ and $\A_{ax}$, respectively. Here the $e_a$, $a = 1, \ldots, \dim G$ are an orthonormal basis for $\g \cong \g^*$ and we sum over repeated indices (so $e_a\otimes e_a$ denotes the identity tensor in $\Sym^2(\g)$).

Evaluating Wilson loop expectations involves two procedures: (i) express the Wilson loop observable $W_{f,\gm}(A)$ as a power series functional in $A$; (ii) insert propagators, each weighted with $\lambda$, into all available slots using the Wick rule. We call each such insertion a \textit{Wick contraction}.

For (i), it suffices to assume $f = \tr \rho$, where $\rho: G \to \mr{End}(V)$ is an irreducible unitary representation of $G$ (since characters form an orthonormal basis in the space of class functions on $G$). As a consequence, we may as well assume $\g$ is embedded inside $\mr{End}(V)$. In this way, we can assume that both $f$ and the inner product $\left<\cdot,\cdot\right>$ on $\g$ are multiples of trace on $\mr{End}(V)$ restricted to $G$ and $\g$, respectively. With these assumptions, we can represent $\hol_\gm(A)$ as a path-ordered exponential that is a power series element in $\mr{End}(V)$, and then apply $f$:
\begin{align}
 W_{f,\gm}(A) &= f(\hol_\gm(A))\\
 &= f(1) + \sum_{n=1}^\infty (-1)^n \int_{1 \geq t_n \geq \ldots \geq t_1 \geq 0}f\Big((\gm^*A)(t_n)\cdots (\gm^*A)(t_1)\Big)\prod_{i=1}^n dt_i. \label{eq:WLexp}
\end{align}

For (ii), when we have a propagator of the general form
$$P = G_{\mu\nu}(u; u')(dx^\mu \otimes dx^\nu) e_a\otimes e_a,$$
where $u, u'$ are points of $\R^2$ and $\mu, \nu = 1,2$, the sum over Wick contractions of (\ref{eq:WLexp}) yields
\begin{align}
\left<W_{f,\gm}\right>_P &= \dim V + \sum_{n=1}^\infty \frac{\lambda^n}{2^nn!}\sum_{\sigma \in S_{2n}}(Lie)_\sigma (An)_\sigma, \label{eq:WLEformula}
\end{align}
where to each permutation $\sigma \in S_{2n}$ we have the corresponding \textit{analytic factor}
\begin{align}
 (An)_\sigma &= \int_{1 > t_{2n} > \ldots > t_1 > 0}\bigg[G_{\mu_{\s(2n)}\mu_{\s(2n-1)}}\Big(\gm(t_{\sigma(2n)}); \gm(t_{\sigma(2n-1)})\Big)\cdots \nonumber \\
 & \hspace{1.5in} G_{\mu_{\s(2)}\mu_{\s(1)}}\Big(\gm(t_{\sigma(2)}); \gm(t_{\sigma(1)})\Big)\bigg]\times \prod_{i=1}^{2n}\dot \gm^{\mu_{\s(i)}}(t_{\s(i)}) dt^i \label{eq:anfactor}
\end{align}
and \textit{Lie factor}
\begin{align}
 (Lie)_\sigma &= \delta_{a_{\s(2n)}a_{\s(2n-1)}}\cdots \delta_{a_{\s(2)}a_{\s(1)}} \tr(e_{a_{2n}}e_{a_{2n-1}}\cdots e_{a_2}e_{a_1}). \label{eq:Liefactor}
\end{align}
As mentioned in the introduction, (\ref{eq:WLEformula}) is a provisional definition, since in (\ref{eq:anfactor}) the Green's functions $G$ are singular. In our case, for $P$ given by $P_{ax}$ and $P_{pax}$, the integrals (\ref{eq:anfactor}) are finite since the singularities in $G$ are tame, but for $P_{ax}$, we get expressions that depend on how we approximate the domain of integration. To avoid singularities along the diagonal of $G$, we took all inequalities $t_{i+1} > t_i$ in (\ref{eq:anfactor}) to be strict.

In computing Wilson loops, we have to specify the class the loops $\gm$ which we are considering. In axial gauge, our (arbitrarily) chosen $y$-direction is a distinguished direction. Thus, it is natural to decompose curves into those which are vertical segments and those which are horizontal curves:

\begin{Definition}\label{Def:horizontal}
A \textit{curve} $\gm$ is a piecewise $C^1$-embedding of an interval $I$ (by default $[0,1]$) into $\R^2$. Since most of our constructions will not depend on how $\gm$ is parametrized, we often conflate $\gm$ with its image in $\R^2$. A curve $\gm: I \to \R^2$ is \textit{horizontal} if it (or its parametrization-reversed curve) is given by $\gm(t) = (t,\bar\gm(t))$ with $\bar\gm(t)$ piecewise $C^1$. We say that a horizontal curve is \textit{right-moving} or \textit{left-moving} if the $x$-coordinate of $\gm$ is increasing or decreasing, respectively.
\end{Definition}


We will often use $x_{\pm}$ to denote the terminal (initial) $x$-coordinates of $\gm$ (so that $x_- < x_+$ if $\gm$ is right-moving and $x_- > x_+$ if $\gm$ is left-moving).

\begin{Definition}
 A curve $\gm$ in the plane is called \textit{admissible} if it can be written as a concatenation $\gm = \gm_m.\cdots.\gm_1$ with each $\gm_i$ a horizontal curve or a vertical segment (we concatenate from right to left). We call such a decomposition an \textit{admissible curve decomposition}. We write
 \begin{equation}
  \gm \equiv \gm_m.\cdots.\gm_1 \label{eq:hor_equiv}
 \end{equation}
 if $\gm$ is the concatenation of the horizontal curves $\gm_i$ up to vertical segments (i.e. we simply omit the vertical segments in an admissible curve decomposition for $\gm$). We say that (\ref{eq:hor_equiv}) is a \textit{horizontal curve decomposition} for $\gm$.
\end{Definition}

In axial gauge, parallel transport along a vertical segment is trivial since the connection has no $dy$ component. Thus, to determine the parallel transport along an admissible curve $\gm$, it suffices to know its horizontal curve decomposition.

Any curve $\gm$ can be approximated in $C^0$ by an admissible curve. Indeed, a piecewise linear approximation $\gm^\sharp$ of $\gm$ will be admissible. In fact, such an approximation can also be chosen so as to make
$$\int_0^1 |\dot\gm(t) - \dot\gm^\sharp(t)|dt$$
arbitrarily small, i.e., $\gm^\sharp$ is ``piecewise $C^1$-close'' to $\gm$. It follows that to prove our main results for arbitrary closed curves, it suffices to establish it for curves which are admissible (or even piecewise linear).\\


\subsection{Generalized White-Noise}

In what follows, we put in quotation marks terms borrowed from probability theory since the intuition they provide is apt. We think of the ($x$-component of) our axial-gauge connection $A^a(x,y)$ as being a Lie-algebra valued ``Gaussian free field'' distributed according to
$$\left<A^a(x,y)A^b(x',y')\right> = \delta^{ab}\lambda G(x,y;y,y'),$$
with $G$ the corresponding Green's function.
Since our Wilson loop operators are obtained by performing parallel transport with respect to the $A^a(x,y)$, our basic ``random variables'' should be obtained by integration of the $A^a(x,y)$ against curves.

\begin{Definition}\label{Def:WN}
 To each horizontal curve $\gm: [x_- ,x_+] \to \R^2$ and $a = 1,\ldots, \dim(G)$, we define ``white-noise'' maps $M^{\gm,a}$ as follows.
 For every $f \in L^2(\R)$, the $M^{\gm,a}(f)$ form a collection a ``Gaussian random variables'', whose covariance is given by
 \begin{align}
\bE\Big(M^{\gm_1, a_1}(f_1)M^{\gm_2, a_2}(f_2)\Big) &= \pm\delta^{a_1a_2} \lambda\int_{x_-^2}^{x_+^2}\int_{x_-^1}^{x_+^1} f_1(x_1)f_2(x_2)G(x_1,\bar\gm_1(x_1); x_2, \bar\gm_2(x_2))dx_2dx_1 \label{eq:gen_cov}
\end{align}
where the $\pm$ is determined by whether the $\gm_i:[x_-^i,x_+^i] \to \R^2$ move in the same direction or opposite direction, respectively. We set
$$M^{\gm} = M^{\gm,a}e_a.$$
to obtain $\g$-valued ``white-noise''. For notational convenience, given an interval $I$, we define
\begin{align*}
M^\gm(I) = M^\gm(\mathbf{1}_I)
\end{align*}
where $\mathbf{1}_I$ is the indicator function of $I$.
\end{Definition}

\begin{Remark}
For the usual (one-dimensional, $\R$-valued) white-noise construction, we replace $\R^2$ with $\R$, let $\gm: \R \to \R$ be the identity map, and let $G(x,x') = \delta(x-x')$ in the above. This yields $\left<M^\gm(f_1), M^\gm(f_2)\right> = \left<f_1,f_2\right>_{L^2}$.
The process $x \mapsto M^\gm([0,x])$ is distributed as Brownian motion.
\end{Remark}

For the partial axial-gauge and complete axial-gauge, we have
 \begin{equation}
  G(x,y; x',y') = \bar G(y,y')\delta(x-x')
 \end{equation}
with $\bar G(y,y')$ the appropriate function of $y$, and so (\ref{eq:gen_cov}) becomes
\begin{equation}
\bE\Big(M^{\gm_1, a}(I_1)M^{\gm_2, b}(I_2)\Big) = \pm\delta^{ab}\lambda\int_{I_1 \cap I_2}\bar G(\bar\gm_1(x),\bar\gm_2(x))dx.
\end{equation}
For $\bar G(y,y')$ given by complete axial-gauge, the induced integral operator is positive definite since $\bar G_{ax}(y,y')$ is the covariance for two independent Brownian motions (moving to the left and right of the origin). However, for $\bar G_{pax}(y,y') = -\frac{1}{2}|y-y'|$ given by partial axial-gauge, we obtain an indefinite operator. Indeed, it is easy to arrange for
\begin{equation}
 \int f(y)dy\int\left(-\frac{1}{2}|y-y'|\right)f(y')dy' \label{eq:Gpairing}
\end{equation}
to be negative since $\bar G_{pax}(y,y') \leq 0$, but if we let $f$ be the sum of two widely spaced bump functions of opposite sign, (\ref{eq:Gpairing}) will be positive.

In this manner, we are led to repeat stochastic analysis with white-noise maps having indefinite covariance, since this is what arises when considering parallel transport operators in partial axial-gauge. However, it turns out that our analysis becomes one of a rather general nature, not just one tied to the specifics of partial axial-gauge. Thus, we develop an abstract framework in the next section that works with general white-noise maps having indefinite covariance. Specializing this abstract framework to the particular white-noise map in Definition \ref{Def:WN}, we prove our main results in Section \ref{Sec:Proofs}.

\section{Algebraic Stochastic Calculus}

Fix a compact Lie algebra $\g \subset \mr{End}(V)$ and a positive integer $m$. Consider ``random variables'' $M^{\al,a}(I)$ indexed by closed intervals $I$ and $\al \in \{0,1, \ldots, m\}$, $a \in \{1, 2, \ldots, \dim \g\}$. We can regard these variables as freely generating an algebra subject to the relation that if $I$ and $J$ have disjoint interior, then
$$M^{\al,a}(I \cup J) = M^{\al,a}(I) + M^{\al,a}(J).$$
In practice, the intervals we consider will always be subintervals of some fixed finite interval. Without loss of generality, we suppose this interval to be $[0,L]$ for some $L > 0$.

\begin{Definition}\label{Def:M}
 Let $C(x) = C^{\alpha\beta}(x)$ be a continuous $(m+1) \times (m+1)$ symmetric-matrix valued function of $x \in [0,L]$, where $0 \leq \al,\beta \leq m$. Let $\lambda$ be a formal variable. The \textit{expectation} operator $\bE$ is the linear functional defined on the algebra generated by the $M^{\alpha,a}(I)$ given by
 \begin{align}
 \bE(M^{\al,a}(I)M^{\beta,b}(J)) &= \delta^{ab}\lambda\int_{I \cap J}C^{\al\beta}(x)dx, \label{eq:algE}
 \end{align}
and which extends to all other monomials by the Wick rule:
\begin{align}
 \bE(X_1\cdots X_n) = \begin{cases}\displaystyle
                       \frac{1}{2^{n/2}(n/2)!}\sum_{\sigma \in S_n}\bE(X_{\sigma(1)}X_{\sigma(2)})\cdots \bE(X_{\sigma(n-1)}X_{\sigma(n)}) & n \textrm{ is even} \\
                       0 & n \textrm{ is odd}.
                      \end{cases}\label{eq:Wick_rule}
\end{align}
The operator $\bE$ extends to the algebra generated by $M^{\al}(I) := M^{\al,a}(I)e_a$ via linearity over $\mr{End}(V)$.\\
\end{Definition}

We are of course interested in the case $C^{\alpha\beta}(x) = \pm\frac{1}{2}|\bar\gm_\al(x)-\bar\gm_\beta(x)|$ relevant to partial axial-gauge, where $\gm_\al$ are a collection of horizontal curves. We ultimately will interpret $\lambda$ as a positive real number, but it is convenient to consider it as a formal parameter in the present general setting to avoid having to deal with convergence issues when we extend $\bE$ to power series elements.

Let $N \geq 1$ be our subdivision (i.e. discretization) parameter, with $N \to \infty$ representing a continuum limit. Our ultimate aim in this section is to prove Lemmas \ref{Lemma:ChangeGauge} and \ref{Lemma:Changevar}. These both strongly rely on the discretization procedure being well-crafted and is the cause for the level of detail in our constructions. Let $\Dx = L/N$ and define the lattice points
$$\Lambda_N = \{i\Delta x: 0 \leq i \leq N\}.$$
This lattice divides $[0,L]$ into the set of subintervals
$$\I_N = \{I_i = [i\Delta x, (i+1)\Delta x]: 0 \leq i \leq N-1\}.$$
These intervals will be used to define discretized integrals (Riemann, It\^{o}, and Stratonovich). Because Stratonovich integrals use the midpoint rule,
we also need to consider $\Dx/2$-translates of these intervals intersected with $[0,L]$. We thus obtain the set of intervals
$$\tilde\I_N = \{J_i = [i\Delta x/2, (i+2)\Delta x/2] : 0 \leq i \leq 2N-2\} \cup \{J_{-1} = [0,\Delta x/2]\}.$$
Given a subinterval $[x_-,x_+] \subset [0,L]$, define
\begin{align*}
\I_{N}[x_-,x_+] &= \{I_i \in \I_N : I_i \subset [x_-,x_+]\} \\
\tilde\I_{N}[x_-,x_+] &= \{J_i \in \tilde \I_N: J_i \subseteq [x_-,x_+]\},
\end{align*}
the set of intervals in $\I_N$ and $\tilde \I_N$ that are contained in $[x_-,x_+]$, respectively.

Let $\M_N$ denote the vector space generated by the basis vectors $M^{\al,a}(I)$ for $I \in \tilde\I_N$, $0 \leq \al \leq m$, $1 \leq a \leq \dim \g$. Then we can regard the $M^\alpha$ as elements of $\M_N \otimes \mr{End}(V).$
Thus, polynomials in the $M^\al$  become elements of $\Sym(\M_N) \otimes \mr{End}(V)$. Let $\widehat{\Sym}(\M_N)$ denote the space of formal power series elements in $\M_N$. So the $M^\al$ belong to the larger space
\begin{equation}
\F_N := \widehat{\Sym}(\M_N) \otimes \mr{End}(V) \label{eq:F_N}
\end{equation}
of power series elements in $\M_N$ with values in $\mr{End}(V)$. The algebra structures on the individual factors in (\ref{eq:F_N}) induce an algebra structure on $\F_N$. Moreover, the natural grading on $\widehat{\Sym}(\M_N)$ (given by polynomial degree) induces one on $\F_N$. Given $f \in \F_N$, write $f^{[i]}$ to denote the degree $i$ part of $f$. It is helpful to think of elements of $f \in \F_N$ as having ``support'' given by the union of the intervals which appear in the terms of $f$.

Given $f \in \F_N$, its expectation $\sum_i \bE(f^{[i]})$ makes sense as a formal power series in $\lambda$, thereby yielding a well-defined map
$$\bE: \F_N \to \mr{End}(V)[[\lambda]].$$
Had we regarded $\lambda$ as a number, the resulting series $\sum_i \bE(f^{[i]})$ may not converge in $\mr{End}(V)$.

We will need to consider inverses of elements in $\F_N$. An element $h \in \widehat{\mr{Sym}}(\M_N) \otimes \mr{End}(V)$ has a multiplicative inverse if and only if its degree zero part $h^{[0]}$ is an invertible element of $\mr{End}(V)$. Given an invertible element $h \in \F_N$, define the corresponding right-adjoint action
\begin{align*}
\ad h:  \F_N & \to \F_N \\
X & \mapsto h^{-1}Xh.
\end{align*}
We have $\tr X = \tr (\mr{ad}(h) X)$ by cyclicity of trace and since $\widehat{\mr{Sym}}(\M_N)$ is a commutative algebra.

Given an $\F_N$-valued function $f$ defined on $\Lambda_N$, we will denote its evaluation at $x \in \Lambda_N$ by $f(x)$ or $f_x$.\\

\begin{Remark}\label{Rem:Functions}
Since $\Lambda_N \subset [0,L]$, a function defined on $[0,L]$ restricts to a function defined on $\Lambda_N$. In the other direction, given a function defined on $\Lambda_N$, it yields a piecewise-constant extension to $[0,L]$ via $f(x) = f(\hat x)$, where $\hat x$ is the largest element of $\Lambda_N$ such that $\hat x \leq x$. In this way, we can pass back and forth between functions defined on $\Lambda_N$ and  $[0,L]$. When we state that a function is defined on $\Lambda_N$, it is to emphasize that it is formed out of objects associated to the discretization paramemter $N$.
\end{Remark}

\begin{Definition}
Fix any $N \geq 1$ and $0 \leq \al \leq m$, and  let $f: [0,L] \to \F_N$ be any function. Then given any interval $[x',x] \subseteq [0,L]$, we can define the following sums:
 \begin{enumerate}
  \item the
  \textit{It\^{o} sum}:
  $${\sum_{[x',x]}}^{(I)_N}f := \sum_{I \in \I_{N}[x',x]} M^\alpha(I)f(x_I^-) $$
where $x_I^-$ is the left endpoint of $I$. This only depends on $f|_{\Lambda_N}$.
 \item the \textit{Stratonovich sum}:
  $${\sum_{[x',x]}}^{(S)_N}f := \sum_{I \in \I_{N}[x',x]}  M^\alpha(I)f(\bar x_I)$$
  where $\bar x_I$ is the midpoint of $I$. This depends only on $f|_{\Lambda_{2N}}$.
  \item \textit{a Riemann sum}:
  $${\sum_{[x',x]}}^{(R)_N, g}f = {\sum_{[x',x]}}^{(R)_N}f := \sum_{I \in \I_{N}[x',x]} g(\tilde x_I^*)|I|f(x_I^*) $$
  where $\tilde x^*_I$ and $x^*_I$ are arbitrary points of $I$ and $g: [0,L] \to \R$ is continuous.
 \end{enumerate}
In the above sums, right instead of left multiplication by the $M^\al(I)$ can be considered as well.
 \end{Definition}

The above Riemann sum is a discrete approximation of $\int f(x)(g(x)dx)$. The above Ito and Stratonivich sums evaluate to elements of $\F_N$ and not ordinary numbers. While we can apply $\bE$ to obtain a number, the resulting numerical sums cannot be estimated using basic tools such as applying the  Cauchy-Schwarz inequality to the pairing (\ref{eq:algE}), since the pairing is not necessarily positive definite. Thus, we have to introduce a bit of terminology in order to organize our estimates.

In ordinary calculus, when we let one of the endpoints of a sum or an integral vary, the result is a new function of the variable endpoint. Hence, in the above, if we let, e.g., the right-endpoint $x$ vary, then we obtain $\F_N$-valued functions on $[0,L]$ (or equivalently, $\F_N$-valued functions on $\Lambda_N$ or $\Lambda_{2N}$ by Remark \ref{Rem:Functions}). In this manner, we can iterate the above sums in the same way we can iterate ordinary sums and integrals.

\begin{Definition}
  Let $f:[0,L] \to \F_N$ be any function. An \textit{admissible sum} is an iterated sum (any combination of It\^{o}, Stratonovich, or Riemann) of $f$
  \begin{equation}
  {\sum_{\mathclap{[x',x]}}}^{(\bullet)_N}\;\;\; {\sum_{\mathclap{[x',x_{n-1}]}}}^{(\bullet)_N}\;\; \cdots \;\; {\sum_{\mathclap{[x',x_1]}}}^{(\bullet)_N} f \label{eq:ad-sum}
  \end{equation}
  where the $k$th sum is with respect to the variable $x_{k-1}$, $2 \leq k \leq n$, and $[x',x]$ is some fixed interval. (Likewise, we can let the left-endpoint $x'$ vary instead of the right, and we can also consider iterated sums using right multiplication). We regard an admissible sum as a sequence of elements $F = (F_N)$, $N \in \mathbb{N}$, with $N$ the subdivision parameter occurring in the definition of the above sums. In this way, an admissible sum is an element of $\prod_{N \geq 1}\F_N$. If all the $F_N \in \F_N$ have degree $i$, i.e. $F_N^{[i]} = F_N$ for all $N$, then we speak of $F$ as having degree $i$. For instance, if $f = f^{[0]}$ in (\ref{eq:ad-sum}), the degree of (\ref{eq:ad-sum}) is the number of It\^{o} and Stratonovich sums occurring in (\ref{eq:ad-sum}). 
\end{Definition}

Admissible sums are essentially a sequence of finer and finer discretizations of a multiple integral. The sense in which we obtain a limiting object will be described below.

Let
$$\mc{X}_N = \{M^\alpha(I): 0 \leq \al \leq m,\; I \in \tilde\I_N\}.$$
We think of the $\X_N$ as infinitesimals, i.e. as differentials, since they are formed out of intervals that are of size $O(N^{-1})$ so that $\bE(X^2) = O(N^{-1})$.

Since each $\F_N$ is an algebra (over $\R$), so is $\prod_{N \geq 1} \F_N$ in the natural way. We will need to estimate expectations of objects which are products of admissible sums and elements of $\mc{X}_N$. Indeed, in the same way that in ordinary calculus, we have $f(x + \Delta x) - f(x) = f'(x)\Delta x + O(\Delta x)^2$, we will need to consider analogous expressions in our algebraic setting. This motivates the following definition:

\begin{Definition}\label{Def:admissible}
  An \textit{admissible monomial} $F$ is an element of $\prod_{N \geq 1} \F_N$ of the form $F = F^{(1)}\cdots F^{(n)}$, where each $F^{(k)} \in \prod_{N \geq 1} \F_N$ is either (i) an admissible sum or else (ii) $F^{(k)} = (F^{(k)}_{N})$ is such that $F^{(k)}_{N} \in \X_N$ for all $N$. An \textit{admissible series} $F$ is an element of $\prod_{N \geq 1} \F_N$ given by a (possibly infinite) linear combination of admissible monomials such that in each polynomial degree $i$, $F^{[i]}$ is a finite linear combination of admissible monomials. 
\end{Definition}

We consider infinite linear combinations of admissible monomials because we need to consider discretized parallel transport operators (see Definitions \ref{Def:PTIto}
and \ref{Def:PTStrat}). Since $\F_N$ is a space of power series elements, our infinite linear combinations are well-defined.

\begin{Definition}\label{Def:order}
Let $F$ be an admissible series. We say $F$ is of \textit{order $O(N^{-k})$} and write
\begin{equation}
F = O(N^{-k})
\end{equation}
if given any admissible series $G_1$ and $G_2$, and any $i \geq 0$, we have
 \begin{equation}
\Big|\bE\Big((G_1FG_2)_N^{[i]}\Big)\Big| \leq CN^{-k} \label{eq:defO}
 \end{equation}
for all $N$, where $C$ is independent of $N$. Similarly, we write
$$F = o(1)$$
if
 \begin{equation}
\lim_{N \to \infty}\Big|\bE\Big((G_1FG_2)_N^{[i]}\Big)\Big| = 0. \label{eq:defo}
 \end{equation}
 
In other words, the above definition prescribes that an admissible series decays (in expectation) at a specified rate in each polynomial degree as we let $N \to \infty$ (i.e. as we refine our partition of $[0,L]$).

We write
\begin{equation}
F = \underline{O}(N^{-k})
\end{equation}
if $F = O(N^{-k})$ and for every $G = O(N^{-k'})$, we have $FG, GF = O(N^{-(k+k')})$. We write $F = O(G_1^{n_1}\cdots G_k^{n_k})$ if $F$ contains the $G_i$ as a multiplicative factors, each with multiplicity at least $n_i$.\\
\end{Definition}

By abuse of notation, we can regard $M^\al(I)$ as an admissible monomial in the sense that we can think of $I$ as placeholder for elements of $\tilde \I_N$, $N \geq 1$. Indeed, in the estimates that follow, we always work at some fixed $N$, but since $N$ is arbitrary, we are in reality choosing suitable $I \in \tilde \I_N$, $N \geq 1$.
In doing so, we also write $F =O(I^n)$ with $I \in \tilde\I_N$ if $F = O(M^{\al_1}(I_1)\cdots M^{\al_n}(I_n))$ with $I_i$ approximately $I$ in the sense that $|I_i| = O(N^{-1})$ and $\mr{dist}(I,I_i) = O(N^{-1}).$

We record some observations. If $F = O(N^{-k})$ with $k > 0$, then
$$\lim_{N\to\infty}\bE(F) = 0.$$
One subtely with the $O(N^{-k})$ notation is that the exponent is not additive under multiplication of differentials:
\begin{align*}
 M^\al(I) &= O(N^{-1})\\
 M^\alpha(I)M^\beta(J) &= O(N^{-1}), \qquad I,J\in \tilde\I_N.
\end{align*}
This is analogous to the case of stochastic differentials, leading to peculiar phenomenon such as that which occurs in It\^{o}'s formula. This is the reason we introduce the $\underline{O}$ notation. However, if $I$ and $J$  are elements of $\tilde \I_N$ with disjoint interior, then
\begin{align*}
 M^\alpha(I)M^\beta(J) &= O(N^{-2}), \qquad I^\circ \cap J^\circ = \emptyset.
\end{align*}

Define variables $M^{\alpha,a}_x$ satisfying
\begin{equation}
\bE(M^{\alpha,a}_x M^{\beta,b}_{x'}) = \delta^{ab}\delta(x-x')C^{\al\beta}(x) \label{eq:EMx}
\end{equation}
Thus,
\begin{equation}
 \int_I M_x^\alpha dx = M^\alpha(I). \label{eq:whitenoise}
\end{equation}
The true meaning of these definitions is that one obtains a well-defined expectation of integrals of the form
\begin{equation}
 I(\vec{\alpha}) := \int_{L > x_n > \ldots > x_1 > 0} M^{\al_n}_{x_n}\cdots M^{\alpha_1}_{x_1}dx_n\cdots dx_1 \qquad \vec{\alpha} = (\alpha_1,\ldots, \al_n), \label{eq:sampint}
\end{equation}
by use of (\ref{eq:EMx}) and the Wick rule. From a more rigorous standpoint, one can regard (\ref{eq:sampint}) as a limit of iterated It\^{o} sums (see Lemma \ref{Lemma:Approx}), each of which is determined by approximating the simplex $\{L > x_n > \ldots > x_1 > 0\}$ with open $n$-cubes and then applying the formula (\ref{eq:whitenoise}). Indeed, consider
\begin{equation}
 I_N(\vec{\alpha}) = \int_{L > x_n >_N \ldots >_N x_1 > 0} M^{\al_n}_{x_n}\cdots M^{\alpha_1}_{x_1} \label{eq:approxint}
\end{equation}
where\footnote{Technically speaking, the condition  $x>_N y$ allows for $x=y$ at the points of $\Lambda_N$ where adjacent intervals meet. But since $\Lambda_N$ is a discrete set, this set of coincidence points is immaterial, i.e. does not contribute to (\ref{eq:sampint}). While this conclusion involves some heuristic reasoning with the formal expression (\ref{eq:sampint}), this conclusion can be seen at the discretized level (\ref{eq:approxint}) by noting that the closed intervals $I_j$ can be replaced by their interior.}
\begin{align}
x >_N y \Leftrightarrow x \in I_i,\, y \in I_j,\, i > j \qquad I_i, I_j \in \I_N.
\end{align}

Then the domain of $I_N(\vec{\al})$ approximates the domain of $I(\vec{\al})$, and $L > x_n >_N \ldots >_N x_1 > 0$ can be written as union of cubes $I_{i_n} \times \cdots\times I_{i_1}$. Applying (\ref{eq:whitenoise}) we can then rewrite $I_N(\vec{\al})$ as the iterated It\^{o} sum
\begin{equation}
 S(\vec{\alpha}) := \sum_{i_n > \ldots > i_1 \atop I_{i_k} \in \I_N}M^{\al_n}(I_{i_n})\cdots M^{\alpha_1}(I_{i_1}). \label{eq:approxIto}
\end{equation}
Note that it was crucial that the inequalities in (\ref{eq:sampint}) were strict in order for it to be approximated by the It\^{o} sums (\ref{eq:approxIto}), since then the limit of the domains of integration of the latter agrees with that of former.

\begin{Lemma}\label{Lemma:Approx}
The limit
$$I(\vec{\al}) := \lim_{N\to\infty}I_N(\vec{\al})$$
exists in the sense that the $I_N(\vec{\al})$ satisfy
$$I_N(\vec{\al}) - I_M(\vec{\al}) = \underline{O}(N^{-1}), \qquad N \leq M.$$
In particular, the expectation of $I(\vec{\al})$ times any admissible series is well-defined. Moreover, given $\vec{\alpha}^{(i)}$ for $1 \leq i \leq m$, then
 \begin{equation}
I(\vec{\alpha}^{(1)})\cdots I(\vec{\alpha}^{(m)}) - I_N(\vec{\alpha}^{(1)})\cdots I_N(\vec{\alpha}^{(m)}) = \underline{O}(N^{-1}).  \label{eq:LemmaApprox}
 \end{equation}
\end{Lemma}

\Proof It suffices to prove the case $m=1$, since
\begin{align*}
I(\vec{\alpha}^{(1)})\cdots I(\vec{\alpha}^{(m)}) -  I_N(\vec{\alpha}^{(1)})\cdots I_N(\vec{\alpha}^{(m)}) &=  \Big(I(\vec{\alpha}^{(1)}) - I_N(\vec{\alpha}^{(1)})\Big)I(\vec{\alpha}^{(2)})\cdots I(\vec{\alpha}^{(m)}) + \ldots \\
& \quad + I_N(\vec{\alpha}^{(1)})\cdots I_N(\vec{\alpha}^{(m-1)})\Big(I(\vec{\alpha}^{(m)}) - I_N(\vec{\alpha}^{(m)})\Big).
\end{align*}

To that end, we first analyze the difference between the domain of integration for $I(\vec{\alpha})$ and that of $I_N(\vec{\alpha})$. It is given by a union of the forbidden regions
$$D_i = \{L > x_n > \cdots > x_1 > 0 : x_{i+1} \not>_N x_i\}.$$
While the path-ordered simplex $\{L > x_n > \ldots x_1 > 0\}$ has fixed volume, the regions $D_i$ have volume $O(N^{-1})$, since they have width $O(N^{-1})$ in the $x_i$--$x_{i+1}$ direction. It is thus enough to replace the left-hand side of (\ref{eq:LemmaApprox}) for $m=1$, with the integrals
\begin{equation}
\sum_{i=1}^n \int_{D_i} M^{\al_n}_{x_n}\cdots M^{\alpha_1}_{x_1}dx_n\cdots dx_1. \label{eq:intoverDi}
\end{equation}
(The $D_i$ are not disjoint, but their overlaps are ``codimension two'', i.e., have volume $O(N^{-2})$ and so will be of lower order.) Technically, we should be considering $I_N(\vec{\al}) - I_M(\vec{\al})
$ instead of (\ref{eq:intoverDi}), but the former's domain of integration will be covered by the $D_i$, so the same proof given below will apply.

If we multiply (\ref{eq:intoverDi}) by any admissible series, the resulting expectation is $O(N^{-1})$ because the volume of the $D_i$ are $O(N^{-1})$. Suppose we multiply (\ref{eq:intoverDi}) by an admissible series $F = O(N^{-k})$, so that it has at least $k$ differentials $M^{\beta_i}(I_j)$ occurring with $|I_j| = O(N^{-1})$. The only way the order of $F$ times (\ref{eq:intoverDi}) can drop below $O(N^{-1-k})$ is if in the sum over Wick contractions occuring between a fixed  $M^{\beta_i}(I_j)$ of $F$ and the terms of (\ref{eq:intoverDi}), we obtain a number that is of order $O(N^{-1})$ instead of $O(N^{-2})$ (i.e. we have two factors of order $O(N^{-1})$ yielding a term of order $O(N^{-1})$). However, it is easy to see that the set of points in $D_i$ which possesses at least one coordinate belonging to a fixed $I_j$ (this is the set where the desired Wick contractions can occur) has volume of order $O(N^{-2})$. Indeed, one must constrain both a coordinate function and the $x_i$--$x_{i+1}$ separation to be $O(N^{-1})$. Thus, the order of $F$ times (\ref{eq:intoverDi}) cannot drop. Thus (\ref{eq:intoverDi}) is $\underline{O}(N^{-1})$.\End


Next, we define discretized versions of the path-ordered exponential. Recall that such a path-ordered exponential represents the solution of an ordinary differential equation, which has the geometric interpretation of parallel transport in the setting of gauge theory. In the (algebraic) stochastic setting, the iterated integrals that appear in the series expansion of parallel transport can be recast as It\^{o} or Stratonovich integrals (sums).

\begin{Definition}\label{Def:PTIto}
Fix $\al$, an initial point $x_-$, and a terminal point $x$. Define \textit{It\^{o} parallel transport} from $x_-$ to $x$ via
$$P^{M^\alpha}_{N,x_{-} \to x} = \begin{cases}\displaystyle
1 + \sum_{n=1}^\infty(-1)^n\sum_{i_n > \ldots > i_1 \atop I_{i_k} \in \I_N[x_{-},x]}M^\alpha(I_{i_n})\cdots M^\alpha(I_{i_1}) & x \geq x_-\\[2ex]
\displaystyle
1 + \sum_{n=1}^\infty\sum_{i_n < \ldots < i_1 \atop I_{i_k} \in \I_N[x,x_-]}M^\alpha(I_{i_n})\cdots M^\alpha(I_{i_1}) & x \leq x_-.
\end{cases}$$
We call these two cases \textit{right-moving} and \textit{left-moving} It\^{o} parallel transport, respectively. In the above, for each $n$ we have an iterated It\^{o} sum since consecutive $I_i$ overlap at their common endpoint.\\
\end{Definition}

Define the relations
\begin{align}
i \succeq j & \Leftrightarrow j = i,\, i - 2,\, i - 4, \ldots \\
i \succ j & \Leftrightarrow j = i - 1,\, i - 3,\, i - 5, \ldots.
\end{align}
and similarly with $\preceq$ and $\prec$.

\begin{Definition}\label{Def:PTStrat}
For $x \geq x_-$, let $i_+$ be the maximum $i$ such that $J_i$ belongs to $\tilde \I_N[x_-,x]$. Likewise, for $x \leq x_-$, let $i_-$ be the minimum $i$ such that $J_i$ belongs to $\tilde \I_N[x,x_-]$. Then define \textit{Stratonovich parallel transport} from $x_-$ to $x_+$ via
$$\tilde P^{M^\alpha}_{N,x_{-} \to x} = \begin{cases}\displaystyle
1 + \sum_{n=1}^\infty(-1)^n\sum_{i_+ \succeq i_n \succ \ldots \succ i_1 \atop J_{i_k} \in \tilde\I_N[x_{-},x]}M^\alpha(J_{i_n})\cdots M^\alpha(J_{i_1}) & x \geq x_-\\[2ex]
\displaystyle
1 + \sum_{n=1}^\infty\sum_{i_- \preceq i_n \prec \ldots \prec i_1 \atop J_{i_k} \in \tilde\I_N[x,x_-]}M^\alpha(J_{i_n})\cdots M^\alpha(J_{i_1}) & x \leq x_-.
\end{cases}$$
We call these two cases \textit{right-moving} and \textit{left-moving} Stratonovich parallel transport, respectively. In the above, for each $n$ we have an iterated Stratonovich sum since consecutive $J_i$ overlap halfway.\\
\end{Definition}

The above parallel transport elements are all admissible series in the sense of Definition \ref{Def:admissible}. 

In stochastic calculus, one can convert Stratonovich integrals to It\^{o} integrals. The same idea allows us to convert from Stratonovich sums to It\^{o} sums up to an error of order $\underline{O}(N^{-1})$. Define
$$i \succ\succ j \Leftrightarrow j = i-2, i-4, \ldots$$

\begin{Lemma} \label{Lemma:StoI} (Stratonovich to It\^{o} conversion)
 We have
\begin{multline}
\sum_{i_+ \succeq i_n \succ \ldots \succ i_1 \atop J_{i_k} \subseteq \tilde\I_N[x_{-},x]}M^{\al_n}(J_{i_n})\cdots M^{\al_1}(J_{i_1}) \\
\begin{split}
&= \sum_{i_+ \succeq i_n \succ\succ i_{n-1} \succ \ldots \succ i_1 \atop J_{i_k} \in \tilde\I_N[x_{-},x]}M^{\al_n}(J_{i_n})\cdots M^{\al_1}(J_{i_1}) \\[1ex]
& \quad + \sum_{\substack{
                  i_+ \succeq i_n \succ\succ i_{n-2} \succ \ldots \succ i_1 \\ i_{n-1} = i_n-1 \\ J_{i_k} \in  \tilde\I_N[x_{-},x]}}
                  \bE\Big(M^{\al_{n}}(J_{i_{n}}) M^{\al_{n-1}}(J_{i_{n-1}})\Big) M^{\al_{n-2}}(J_{i_{n-2}})\cdots M^{\al_1}(J_{i_1})\\[1ex]
& \qquad + \underline{O}(N^{-1})
\end{split}\label{eq:StoIto}
\end{multline}
and similarly when iterating sums from left to right.
\end{Lemma}
In other words, in going from the left-hand side to the right-hand side in the above equation, the Stratonovich sum over the $i_n$ index was converted an It\^{o} sum at the expense of an expectation of the $i_n$ and $i_{n-1}$ terms (which is nonzero only if $i_{n-1} = i_n - 1$, i.e., if $J_{i_{n-1}}$ and $J_{i_n}$ overlap) and an error of order $\underline{O}(N^{-1})$. Iterating this procedure we can convert all Stratonovich sums into It\^{o}--Riemann sums up to $\underline{O}(N^{-1})$.\\

\Proof We need to show that
\begin{equation}
\sum_{i_+ \succeq i_n \atop J_{i_n} \in \tilde\I_N[x_-,x]}\left[M^{\al_n}(J_{i_{n}})M^{\al_{n-1}}(J_{i_n - 1}) - \bE\Big(M^{\al_n}(J_{i_{n}})M^{\al_{n-1}}(J_{i_n - 1})\Big)\right] \label{eq:M2}
\end{equation}
is $\underline{O}(N^{-1})$. First, each term of the sum of (\ref{eq:M2}) is $O(N^{-2})$ since it is quadratic in the $M^\al(J)$ and has zero expectation. On the other hand, we have a sum over $O(N)$ many terms. The same analysis in the proof of Lemma \ref{Lemma:Approx} shows that we obtain a result that is $\underline{O}(N^{-1})$, since the order of (\ref{eq:M2}) cannot drop when multiplying by a differential (the terms of (\ref{eq:M2}) have supports that are ``sparse'', so that if we wish to consider those that have support on an interval of length $O(N^{-1})$, we pick up only finitely many terms).\End

Given an interval $J = [x,x+\Delta x] \in \tilde\I_N$, write
\begin{align*}
J^{-} &= [x,x+\Delta x/2] \\
J^{+} &= [x+\Delta x/2, x + \Delta x].
\end{align*}
to denote the subintervals obtained from division of $J$ along its midpoint. For the special interval $J_{-1} \in \tilde \I_N$ of length $\Delta x/2$,  define
\begin{align*}
J^{-}_{-1} &= \{0\} \\
J^{+}_{-1} &= J_{-1}.
\end{align*}
These intervals belong to $\I_{2N}$. They satisfy
\begin{align}
 J^{+}_i &= [(i+1)\Dx/2, (i+2)\Dx/2], \qquad -1 \leq i \leq 2N-2 \label{eq:Ji+}\\
 J_i^{+} &= J_{i+1}^{-}
\end{align}
so that the $J^+_i$ partition $[0,L]$ and intersect only at common endpoints.

Write $(f \circ M)(J)$ to mean $f(\bar x)M(J)$ where $\bar x$ is the midpoint of $J$ (this assumes that for $f$ defined on a lattice, $\bar x$ is a corresponding lattice point). Define
\begin{equation}
(f \hat\circ M)(J) =
(f \circ M)(J^{-}) + (f \circ M)(J^{+})
\end{equation}

From our $M^\al$, $0 \leq \al \leq m$, we isolate $\al = 0$ and define
\begin{equation}
 h_{N,x} = P^{M^0}_{N,0 \to x}, \qquad x \in \Lambda_N.
\end{equation}
We think of $h_N$ as a gauge-transformation defined on the lattice site $x$. It satisfies a discretized It\^{o} differential equation:
$$h_{N,x+\Dx} - h_{N,x} = -M^0([x,x+\Dx])h_{N,x}.$$

\begin{Lemma}
 Let $J \in \tilde \I_N$. For $x \in \Lambda_{4N}$, let $I_x = [x,x+\Dx/4]$. We have
 \begin{align*}
\ad (h_{4N,x + \Dx/4})M^\alpha(J) &= \ad (h_{4N,x})M^\alpha(J) + O([M^0(I_x), M^\alpha(J)]) + O(M^0(I_x)^2, M^\alpha(J))\\
 &= \ad (h_{4N,x})M^\alpha(J) + O(N^{-2}).
 \end{align*}
\end{Lemma}

\Proof For any increment $\Delta x$, we have
\begin{align*}
\Big (\ad (h_{4N,x+\Delta x}) - \ad (h_{4N,x})\Big)X &= \Big(h_{4N,x+\Delta x}^{-1} - h_{4N,x}^{-1}\Big)X h_{4N,x} + h^{-1}_{4N,x}X\Big(h_{4N,x+\Delta x}-h_{4N,x}\Big) + \\
 &\qquad + \Big(h_{4N,x+\Delta x}^{-1} - h_{4N,x}^{-1}\Big)X (h_{4N,x+\Delta x} - h_{4N,x})
\end{align*}
Now, when we do a single increment $\Delta x/4$ for $h_{4N,x}$ we get
\begin{align*}
h_{4N,x + \Dx/4} &= (1 - M^0(I_x))h_{4N,x} \\
h_{4N,x + \Dx/4}^{-1} &= h_{4N,x}^{-1}(1 - M^0(I_x))^{-1}\\
&= h_{4N,x}^{-1}\big(1 + M^0(I_x) + \ldots\big).
\end{align*}
Hence,
 \begin{align}\begin{split}
 \Big(\ad (h_{4N,x + \Dx/4}) - \ad (h_{4N,x})\Big)M^\alpha(J) &= \ad (h_{4N,x})[M^0(I_x),\, M^\alpha(J)]\\
 &\qquad + O(M^0(I_x)^2, M^\alpha(J)).\end{split} \label{eq:aderror}
\end{align}
Now,
\begin{align*}
\bE([M^0(I_x), M^\alpha(J)]) &= [e^a,e^b]\bE\Big(M^{0,a}(I_x)M^{\alpha,b}(J)\Big) \\
&= 0.
\end{align*}
since the expectation is nonvanishing only for $a=b$, in which case $[e^a,e^a]=0$. So the right-hand side of (\ref{eq:aderror}) is $O(N^{-2})$.\End

\begin{Lemma}\label{Lemma:split}
Let $J \in \tilde\I_N$. We have
 \begin{align*}
  \mr{ad}(h_{2N}) \circ M^\alpha(J) &=
  \mr{ad}(h_{4N}) \hat\circ M^\alpha(J) + O(N^{-2})
 \end{align*}
 with the remainder $O(N^{-2})$ consisting of terms of the form $O(J^2)$ and $\underline{O}(N^{-1})O(J)$.
\end{Lemma}

\Proof For $J = [x,x+\Dx]$, we have
\begin{align*}
 \mr{ad}(h_{4N}) \hat\circ M^\alpha(J) &= \ad (h_{4N,x + \Dx/4})M^\alpha(J^-) + \ad (h_{4N,x+3\Delta x/4})M^\alpha(J^+)\\
 \mr{ad}(h_{2N}) \circ M^\alpha(J) &= \mr{ad}(h_{2N,x+\Delta x/2})M^\alpha(J^-) + \mr{ad}(h_{2N,x+\Delta x/2})M^\alpha(J^+)\\
 &= \mr{ad}(h_{4N,x+\Delta x/2})M^\alpha(J^-) + \mr{ad}(h_{4N,x+\Delta x/2})M^\alpha(J^+) + \underline{O}(N^{-1})O(J)
\end{align*}
where we used Lemma \ref{Lemma:Approx} in the last line. Now apply the previous lemma.\End

Let $x^\al_\pm$ be points in $[0,L]$, $1 \leq \al \leq m$, satisfying the matching conditions
\begin{equation}
x^1_+ = x^2_-, \quad x^2_+ = x^3_-,\quad \ldots, \quad x^m_+ = x^1_-. \label{eq:match}
\end{equation}
Define the function $g^\al_N$ on $\Lambda_N$ using either of the following choices
\begin{align}
g^\al_{N,x} = \begin{cases}
P^{M^\al}_{N,x_-^\al \to x} & \textrm{only if }C^{\al\al} \equiv 0\\
         \tilde P^{M^\al}_{N,x_-^\al \to x} &  \textrm{general }C^{\al\beta}.
        \end{cases} \label{eq:defg}
\end{align}

We want to compute
\begin{equation}
\lim_{N \to \infty}\bE\Big(\tr (g^m_{N,x_+^m}\cdots g^1_{N,x_+^1})\Big). \label{eq:discWLE}
\end{equation}
For $C^{\al\beta}$ pertaining to partial or complete axial-gauge, (\ref{eq:discWLE}) is precisely our corresponding Wilson loop expectation. We can rewrite the trace occuring in (\ref{eq:discWLE}) as
\begin{align*}
\tr \Big(h^{-1}_{N,x^m_+} g^{m}_{N,x^m_+}h_{N,x^m_-} h^{-1}_{N,x^{m-1}_+}g^{m-1}_{N,x_+^{m-1}}h_{N,x^{m-1}_-}\cdots h^{-1}_{N,x^1_+}g^{1}_{N,x^1_+}h_{N,x_-^1}\Big).
\end{align*}
due to the matching condition (\ref{eq:match}).

So consider
\begin{equation}
 \hat g_{N,x}^\al = h_{2N,x}^{-1}g_{2N,x}^\al h_{2N,x^\al_-}, \qquad x \in \Lambda_{2N}. \label{eq:hatg}
\end{equation}
The use of $2N$ on the right-hand side of (\ref{eq:hatg}) is to faciliate analysis at midpoints needed when working with Stratonovich sums. If we regard $g^\al_{2N,x}$ as parallel transport induced by $M^\al$, we are to regard $\hat g^\al_{N,x}$ as parallel transport induced by the gauge-transform of $M^\al$ by $h_{2N}$.

Define
\begin{equation}
\tilde M^\alpha(I) = M^\alpha(I) - M^0(I),\qquad \alpha = 1,\ldots, m. \label{def:tildeM}
    \end{equation}
These variables satisfy
\begin{equation}
 \bE(\tilde M^\al(I),\tilde M^\beta(J)) = \int_{I \cap J}\tilde C^{\al\beta}(x)dx \label{eq:EtildeM}
\end{equation}
where
\begin{equation}
 \tilde C^{\al\beta}(x) = C^{\al\beta}(x) - C^{\al 0}(x) - C^{0\beta}(x) + C^{00}(x). \label{eq:tildeC}
\end{equation}

\begin{Lemma}\label{Lemma:ChangeGauge} (Change of gauge)
 Suppose $g^\al_{2N,x}$ is right-moving. Then
 \begin{align}
  \hat g_{N,x}^\alpha = 1+\sum_{n=1}^\infty(-1)^n\sum_{i_+ \succeq i_n \succ \ldots \succ i_1 \atop J_{i_k}\in \tilde\I_N[x_-^\alpha, x]} (\mr{ad}(h_{4N}) \hat\circ \tilde M^\alpha)(J_{i_n})\cdots (\mr{ad}(h_{4N}) \hat\circ \tilde M^\alpha)(J_{i_1}) + o(1) \label{eq:tildegdecomp}
 \end{align}
 The analogous result holds for $g^\al_{N,x}$ left-moving.
\end{Lemma}

The expression (\ref{eq:tildegdecomp}) is an admissible series formed out of iterated Stratonovich sums. Because $\bE(\tilde M^\al(I), \tilde M^\al(J)) \neq 0$ for overlapping intervals, these Stratonovich sums remain distinct from iterated It\^{o} sums as $N \to \infty$.\\

\Proof We consider the case when $g^\al_N$ is defined by the first case of (\ref{eq:defg}), with the second case following similar lines. Let $x_i = i\Dx$, $0 \leq i \leq N$ and $\bar x_i$ the midpoint of $[x_i,x_{i+1}]$. For $f$ defined on $\Lambda_{2N}$, define the forward difference operators
\begin{align*}
\Delta_i f &= f_{x_{i+1}} - f_{x_i}\\
\bar\Delta_i f &= f_{\bar x_i} - f_{x_i}, \qquad 0\leq i\leq N-1.
\end{align*}

For $x = x_k$, we have
$$f_x = \sum_{i=0}^{k-1} \Delta_i f.$$
and for any two functions $f$ and $g$, we have
\begin{align}
\Delta_i(fg) &= (\Delta_if)g_{x_i} + f_{x_i}\Delta_ig + \Delta_if\Delta_ig \label{eq:proddiff1}\\
&= (\Delta_if)g_{\bar x_i} + f_{\bar x_i} \Delta_i g + \Big(\Delta_if\Delta_ig - (\bar\Delta_if)\Delta_ig - \Delta_if\bar\Delta_ig\Big) \label{eq:proddiff2}
\end{align}
Equation (\ref{eq:proddiff1}) is an It\^{o}-type formula for the differential of a product. Equation (\ref{eq:proddiff2}), which uses the midpoint rule, converts the It\^{o} differentials in (\ref{eq:proddiff1}) into Stratonovich differentials.

We have
\begin{align*}
\bar\Delta_i g_{2N}^\alpha &= -M^\alpha([x_i, \bar x_i])g^\al_{2N,x_i}\\
 \Delta_i g_{2N}^\alpha &= g_{2N,x_{i+1}}^\al - g_{2N,\bar x_{i}}^\al + g_{2N,\bar x_{i}}^\al - g_{2N,x_{i}}^\al\\
 &= \Big(-M^\alpha([\bar x_i, x_{i+1}])(-M^\al([x_i, \bar x_{i}]) + 1) - M^\al([x_i,\bar x_i])\Big)g^\al_{2N,x_i}\\
 &= \Big(-M^\alpha([x_i, x_{i+1}]) + M^\alpha([\bar x_i, x_{i+1}])M^\alpha([x_i,\bar x_i])\Big)g^\al_{2N,x_i}
 \end{align*}
and similarly with $\Delta_i h_{2N}$ and $\bar\Delta_ih_{2N}$. Thus,
\begin{align*}
\bar\Delta_i h^{-1}_{2N} &= h_{2N,x_i}^{-1}\Big(\Big(1 - M^0([x_i,\bar x_i])\Big)^{-1} - 1\Big)\\
\Delta_ih_{2N}^{-1}  &= h_{2N,x_i}^{-1} \Big(\Big(1 - M^0([x_i,x_{i+1}]) + M^0([\bar x_i,x_{i+1}])M^0([x_i,\bar x_i])\Big)^{-1}-1\Big).
\end{align*}
For notational clarity, we temporarily drop the superscript $\alpha$ on the $g^\al_{2N}$ and $\hat g^\al_N$ below.

Making use of  (\ref{eq:proddiff2}) we have
\begin{align}
 \Delta_i \hat g_N &=
 \Big((\Delta_i h^{-1}_{2N})g_{2N,\bar x_i} + h^{-1}_{2N,\bar x_i}(\Delta_i g_{2N})\Big)h_{2N,x^\alpha_-} + R_i \label{eq:tildeg1}
 \end{align}
where the remainder $R_i$ is given by
\begin{align}
R_i
&= h_{2N,x_i}^{-1}\Big(-M^0([x_i,x_{i+1}])M^\alpha([x_i,x_{i+1}]) + M^0([x_i,\bar x_i])M^\al([x_i,x_{i+1}]) + \nonumber\\
& \qquad + M^0([x_i,x_{i+1}])M^\al([x_i,\bar x_i]) + O([x_i,x_{i+1}]^3)\Big)g_{2N,x_i}h_{2N,x_-^\al}.\label{eq:rem}
\end{align}
The expectation of the terms quadratic in the $M$'s is equal to
\begin{align*}
-\int_{[x_i,x_{i+1}]}C^{0\al}(x)dx + 2\int_{[x_i,\bar x_i]}C^{0\al}(x)dx = \int_{[x_i,\bar x_i]}[C^{0\al}(x) - C^{0\al}(x+\Delta x/2)]dx
\end{align*}
times $\sum_{a} e_ae_a \in \mr{End}(V)$. Since $C$ is continuous, then $C^{0\al}(x) - C^{0\al}(x+\Delta x/2) = o(1)$ as $N \to \infty$. It follows that $R_i$ is asymptotically of the form $o(1)\underline{O}(N^{-1}) + O(N^{-2}) = o(1)O(\underline{N}^{-1})$. In what follows the precise nature of the remainder $R_i$ may change from line to line; it is only required to be a function that is of the form of the form $O([x_i,x_{i+1}]^2)$ and asymptotically $o(1)\underline{O}({N}^{-1})$.

Consider the leading term of (\ref{eq:tildeg1}). We want to write  $\Delta_i h_{2N}^{-1}$ and $\Delta_i g_{2N}$ in terms of the midpoint of the corresponding function in order to get differentials that are of Stratonovich type. Thus,
\begin{align*}
\Delta_i g_{2N} &= g_{2N,x_{i+1}} - g_{2N,\bar x_{i}} + g_{2N,\bar x_{i}} - g_{2N,x_{i}}\\
&= -M^\al([\bar x_i, x_{i+1}])g_{2N,\bar x_i} + \Big(1 - \Big(1-M^\al([x_i,\bar x_i])\Big)^{-1}\Big)g_{2N,\bar x_i} \\
&= -M^\al([x_i,x_{i+1}])g_{2N,\bar x_i} + R_i\\
\Delta_i h_{2N}^{-1} &= h_{2N,x_{i+1}}^{-1} - h_{2N,\bar x_{i}}^{-1} + h_{2N,\bar x_{i}}^{-1} - h_{2N,x_{i}}^{-1}\\
&= h_{2N,\bar x_i}^{-1}\Big(\Big(1 - M^0([\bar x_i,x_{i+1}])\Big)^{-1}-1\Big) + h_{2N,\bar x_i}^{-1}\Big(1 - \Big(1 - M^0([x_i,\bar x_i])\Big)\Big)\\
&= h_{2N,\bar x_i}^{-1}M^0([x_i,x_{i+1}]) + R_i.
\end{align*}
Here, we used the fact that terms quadratic in $M^\alpha$ or $M^0$ are $O(N^{-2})$, due to the isotropic condition $C^{\alpha\alpha} = C^{00} = 0$, so that we may shuffle them into the remainder $R_i$.

Thus (\ref{eq:tildeg1}) becomes
\begin{align}
 \Delta_i \hat g_N &= h_{2N,\bar x_i}^{-1}\Big(M^0([x_i,x_{i+1}]) - M^\alpha([x_i,x_{i+1}])\Big)g_{2N,\bar x_i}h_{2N,x_-^\al} + R_i \nonumber \\
 &= -h_{2N,\bar x_i}^{-1}\tilde M^\al([x_i,x_{i+1}])h_{2N,\bar x_i}(h_{2N,\bar x_i}^{-1}g_{2N,\bar x_i}h_{2N,x_-^\al}) + R_i \nonumber \\
 &= -\Big(\big(\mr{ad}(h_{2N}) \circ \tilde M^\al\big)([x_i,x_{i+1}])\Big)\hat g_{N,\bar x_i} + R_i \nonumber \\
 &= -\Big(\big(\mr{ad}(h_{4N}) \hat\circ \tilde M^\al\big)([x_i,x_{i+1}])\Big)\hat g_{N,\bar x_i} + R_i \label{eq:discSDE}
 \end{align}
where we used Lemma \ref{Lemma:split} in the last line.

Equation (\ref{eq:discSDE}) says $\hat g_N$ satisfies a discretized Stratonovich differential equation up to a lower order remainder $R_i$. Since terms of order $o(1)\underline{O}(N^{-1})$ are subdominant when looking at order $\Dx$ (the order of a single increment), we expect $\hat g_{N,x}$ to be close to
\begin{align}
  g_{N,x}^{\sharp} := 1+\sum_{n=1}^\infty(-1)^n\sum_{i_+ \succeq i_n \succ \ldots \succ i_1 \atop I_{i_k} \in \tilde\I_{N}[x_-^\al,x]} (\mr{ad}(h_{4N}) \hat\circ \tilde M^\alpha)(J_{i_n})\cdots (\mr{ad}(h_{4N}) \hat\circ \tilde M^\alpha)(J_{i_1}),
\end{align}
which is a discretized path-ordered exponential whose individual terms are iterated Stratonovich sums. Since $g^{\sharp}_{N,x}$ is given by iterated Stratonovich sums, its increment satisfies
$$\Delta_i g^{\sharp}_N = -(\ad(h_{4N})\hat\circ\tilde M^\al)([x_i,x_{i+1}])g^{\sharp}_{N,\bar x_i},$$
a midpoint rule rather than a left endpoint rule.

We have
\begin{align}
 \hat g_{N,x_{i+1}} - g^{\sharp}_{N,x_{i+1}} &=
 (\hat g_{N,x_{i+1}} - \hat g_{N,x_i}) + (\hat g_{N,x_i} - g^{\sharp}_{N,x_i}) - (g^{\sharp}_{N,x_{i+1}} - g^{\sharp}_{N,x_i}) \nonumber \\
 &= \Big(-(\mr{ad}(h_{4N})\hat\circ\tilde M^\al)([x_i,x_{i+1}]) \hat g_{N,\bar x_i} + R_{i}\Big) \nonumber \\
 & \qquad + (\hat g_{N,x_{i}} - g^{\sharp}_{N,x_i}) + \Big((\mr{ad}(h_{4N})\hat\circ\tilde M^\al)([x_i,x_{i+1}])\Big)g^{\sharp}_{N,\bar x_i} \nonumber\\
\begin{split}
 &= \Big(-(\ad(h_{4N})\hat\circ \tilde M^\al)([x_i,x_{i+1}])\Big)\big(\hat g_{N, \bar x_i}-g^{\sharp}_{N, \bar x_i}\big)\\
 &\qquad +  (\hat g_{N,x_i}- g^{\sharp}_{N,x_i}) + R_i. \label{eq:error}
\end{split}
\end{align}
Instead of incrementing from $x_i$ to $x_{i+1}$ by step size $\Delta x$, we could have also incremented from $\bar x_i$ to $\bar x_{i+1}$. In doing so, the above analysis can be repeated to show that
\begin{align}
 \begin{split}
  \hat g_{N,\bar x_{i+1}} - g^{\sharp}_{N, \bar x_{i+1}} &= \Big(-(\ad(h_{4N})\hat\circ \tilde M^\al)([\bar x_i, \bar x_{i+1}])\Big)\big(\hat g_{N, x_{i+1}}-g^{\sharp}_{N,  x_{i+1}}\big)\\
 &\qquad +  (\hat g_{N,\bar x_i}- g^{\sharp}_{N,\bar x_i}) + R_i. \label{eq:error2}
 \end{split}
\end{align}

Let $y_j = j\Delta x/2$ be an enumeration of the points $x_i$, $\bar x_i$ of $\Lambda_{2N}$, $j = 0, \ldots, 2N$. Letting
\begin{align}
 a_j &= \hat g_{N,y_j} -  g^{\sharp}_{N,y_j}, \\
 c_j &= -(\ad(h_{4N})\hat\circ \tilde M^\al)([y_{j-1}, y_{j+1}]),
\end{align}
then (\ref{eq:error}) and (\ref{eq:error2}) imply that we have the recurrence relation
\begin{equation}
 a_{j+2} = c_{j+1}a_{j+1} + a_j + r_{j+1}
\end{equation}
where $r_j$ denotes a remainder that is of the form $O([y_{j-1},y_{j+1}]^2)$ and asymptotically $o(1)\underline{O}(N^{-1})$. Solving this recurrence relation with
\begin{align*}
 a_0 &= 0\\
 a_1 &= r_0
\end{align*}
we find that
\begin{align}
 a_j = \sum_{\bar j=0}^{j-1}P_{\bar j}^{(j)}r_{\bar j} \label{eq:a_j}
\end{align}
where $P^{(j)}_{\bar j}$ is a polynomial in $c_1,\ldots, c_j$ such that
\begin{itemize}
 \item $\deg P^{(j)}_{\bar j} = j-1-\bar j$;
 \item individual monomial terms consist of products of distinct $c_i$'s, whose indices occur in descending order from left to right.
\end{itemize}

We want to show that $a_j = \hat g_{N,y_j} -  g^{\sharp}_{N,y_j}$ is $o(1)$ for all $0 \leq j \leq 2N$. It suffices (and it is imperative) that we work at some fixed order in powers of $\lambda$ as we let $N \to \infty$, since the presence of $h^{-1}_{4N}$ terms occurring in $\ad (h_{4N})$ forbid us to work at all orders uniformly in $N$. Thus, we need only consider terms of (\ref{eq:a_j}) that are at most of polynomial degree $\ell$ in the $c_j$'s, for arbitrary fixed $\ell$.  In what follows, estimates are uniform with respect to fixed $\ell$.

For $d \leq \ell$, the number of (monic) monomials of degree $d$ in the $c_j$ is at most $O(N^d)$, since $j$ ranges from $1, \ldots, 2N-1$. A generic term of the form
\begin{equation}
c_{j_d}\cdots c_{j_1}r_{j_0}, \qquad j_d > \ldots > j_0 \label{eq:monomial}
\end{equation}
is of order $o(1)O(N^{-(d+1)})$. This occurs when $j_{k+1} - j_k > 1$ for all $k = 0, \ldots, d-1$ so that the support of the infinitesimals occuring in the $c_{j_k}$ and $r_{j_0}$ are disjoint. For each adjacency $j_{k+1} = j_k + 1$, the order of (\ref{eq:monomial}) increases by a factor of $N$, but then the number of monomials with this condition also decreases by a factor of $O(N)$. So for each fixed $\bar j$ in (\ref{eq:a_j}), considering all the monomials of $P_{\bar j}^{(j)}$ of degree at most $\ell$, we obtain an overall term that is $\ell\cdot o(1)O(N^{-1})$. Indeed, we have (essentially) a sum of $O(N^d)$ terms of order $o(1)O(N^{-(d+1)})$, for $d = 0, 1, \ldots, \ell$. Summing over $\bar j$, we obtain $O(N)$ terms of order $\ell\cdot o(1)O(N^{-1})$, so that $a_j$ is $o(1)$ for any fixed $\ell$. Since $\ell$ is arbitrary, this shows that $\hat g_N = g^\sharp_N + o(1)$.\End

\begin{figure}
\includegraphics{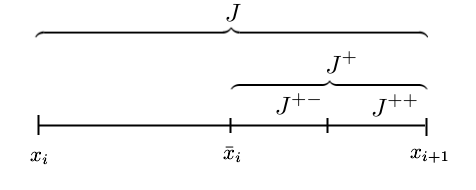}
\caption{Nesting of intervals when $J = [x_i,x_{i+1}]$.}
\end{figure}

Let $\tilde \I_N^+ = \{J^+: J \in \tilde\I_N\}$. From (\ref{eq:Ji+}), we have
\begin{align}
\tilde \I^+_N = \I_{2N}.
\end{align}
For $J \in \tilde \I_N^+$, define
\begin{align}
J^{++} := (J^+)^+\\
J^{+-} := (J^+)^-.
\end{align}
These intervals are of size $\Delta x/4$ and partition $[0,L]$.
Call the resulting set of intervals $\tilde\I_N^{++}$ and $\tilde\I_N^{+-}$, respectively. So we have
\begin{equation}
\I_{4N} = \tilde \I_N^{+-} \cup \tilde \I_N^{++}.
\end{equation}

Observe that $h_{4N}$ is an iterated It\^{o} sum formed out of intervals belonging to $\I_{4N}$. For the next lemma, we make crucial use that $h$ is defined on $\Lambda_{4N}$ using It\^{o} sums. (By comparison, in Lemma \ref{Lemma:ChangeGauge}, $h$, like the $g^\al$, could have been defined using the Stratonovich integral as in the second case of (\ref{eq:defg}).

 \begin{Lemma}\label{Lemma:Changevar} (Gauge-invariance)
 Let $P$ be any polynomial in the $M^{\al}(J)$, with $J \in \tilde\I_N$, $0 \leq \al \leq m$. Regarding $P$ as the induced polynomial in the $M^{\al}(J^+)$ as well, $J^+ \in \tilde \I_N^+$,  let $P^h$ be the corresponding polynomial with each $M^{\al}(J^+)$ replaced with $(\ad(h_{4N})\circ M^{\al})(J^+)$. Then
 \begin{equation}
\bE\big(\tr\, P\big) = \bE\big(\tr\, P^h\big) \label{eq:findimexp}
 \end{equation}
\end{Lemma}

\Proof Let $V_N$ be the (real) vector space spanned by a basis consisting of elements that are in one-to-one correspondence with the set
\begin{equation}
\mc{V}_N = \{M^{\alpha,a}(I) : 1 \leq \alpha \leq m,\; I \in \I_{2N}\} \cup \{M^{0,a}(\tilde I) : \tilde I \in \I_{4N}\}. \label{eq:variables}
\end{equation}
Thus, the elements of $\V_N$ can be regarded as coordinate functions, i.e., linear functionals, on $V_N$. The expectation operator restricted to polynomial functions on $V_N$ (i.e. polynomials in the variables appearing in $\mc{V}_N$) is completely determined by the covariance matrix
$$\C = \C_{(\al,I,a), (\beta,J,b)} := \bE(M^{\al,a}(I)M^{\beta,b}(J)), \qquad M^{\al,a}(I), M^{\beta,b}(J) \in \mc{V}_N.$$
Observe that $\C$ is a tensor product of two matrices: one parametrizing the $0 \leq \al \leq m$ index and interval variables $I$ and the other the identity matrix with respect to Lie-algebra indices. In what follows, all our matrices factorize in this way into a non Lie-algebra part tensored the identity matrix on the Lie algebra.

The matrix $\C$ is not necessarily invertible, but by adding an arbitrarily small matrix, call it $\eps$, we can make $\C + \eps$ is invertible (since $\C$ is weighted by $\lambda$, we do the same for $\eps$). Let $A_\eps$ be its inverse. Define $\bW_{A_\eps}$ to be the operator on power series functions on $V_N$ given by applying the Wick rule (i.e. the rule (\ref{eq:Wick_rule})) using the covariance $\C + \eps$. If $A+\eps$ were positive-definite, then $\bW_{A_\eps}$ could be expressed as integration against a Gaussian measure
$$d\mu_{A_\eps} = c_{A_\eps} e^{-(v,A_\eps v)/2}\prod_{X \in \V_N} dX,$$
where $c_{A_\eps}$ is a normalization constant and $(v,A_\eps v)$ stands for the pairing on $V_N$ dual to the pairing $\C+\eps$ on $V_N^*$ (which in the appropriate basis is given by the inverse matrix $A_\eps$ to $\C + \eps$). In other words,
\begin{equation}
\bW_{A_\eps}(f) = \int_{V_N} f(v)d\mu_{A_\eps}(v). \label{eq:WickExp}
\end{equation}
Thus, for $\C$ positive-definite (and hence also $\C + \eps$ for $\eps$ small) the expression on the right-hand side of (\ref{eq:WickExp}), being an integral, is invariant under changes of coordinates via the usual properties of integrals. However, in \cite{Ngu-PI}, it is shown that the series in $\lambda$ one obtains on the left-hand side of (\ref{eq:WickExp}) using the Wick rule, i.e., the \textit{Wick expansion}, is invariant under changes of coordinates. In other words, while we no longer have a Gaussian measure in general for the right-hand side of (\ref{eq:WickExp}), formal calculus manipulations still hold in the sense that if $\Theta: V_N \to V_N$ is a diffeomorphism (or more generally, an invertible power series) that fixes the origin, then writing
$$\Theta^*\Big(f(v)d\mu_{A_\eps}(v)\Big) = \tilde f(v)d\mu_{\tilde A}(v)$$
where the right-hand side is expressed using the usual change of variables formula, we have
\begin{equation}
\bW_{A_\eps}(f) = \bW_{\tilde A}(\tilde f). \label{eq:equalWick}
\end{equation}
In short, (\ref{eq:equalWick}) is a well-defined algebraic identity between two formal series in $\lambda$, without regard to there being an honest measure (though the expressions for $\tilde A$ and $\tilde f$ are obtained as though one were doing a change of variables for a measure.) See \cite[Theorem 1.5]{Ngu-PI} for a proof of (\ref{eq:equalWick}). See also \cite{Ngu_video1} for additional details.

We prove (\ref{eq:findimexp}) from (\ref{eq:equalWick}) by showing that the right-hand side is a change of coordinates in a Wick expansion, and hence, the result is unaffected. So consider the power series change of variables
\begin{align*}
 \Theta: V_N & \to \widehat{\Sym}(V_N) \\
 X & \mapsto \ad(h_{4N,x^*})X
\end{align*}
where if $X = M^\al(J)$, then
\begin{align*}
 x^* = \begin{cases}
        \textrm{midpoint} & J \in \tilde \I_N^+ = \I_{2N}\\
        \textrm{right endpoint} & J \in \tilde \I_N^{+-}\\
        \textrm{left endpoint} & J \in \tilde\I_N^{++}.
       \end{cases}
\end{align*}
This choice of $x^*$ is so that if $I$ and $J$ are intervals from $\tilde\I_N^+ \cup \tilde\I_N^{+-} \cup \tilde\I_N^{++}$ that have overlapping interior, the corresponding $x^*$ are equal. This is crucial in what follows and is ultimately the reason why our discretizions and interval subdivisions are as they are.

So we have
\begin{equation}
\int_{V_N} \tr (P)d\mu_{A_\eps} = \int_{V_N}\Theta^*\Big(\tr (P)d\mu_{A_\eps}\Big). \label{eq:WickExp2}
\end{equation}
in the sense of Wick expansions. Now, $\Theta^*\tr (P) = \tr (P^h)$. Here, we used that
$$(\ad h_{4N} \circ M^\al)(J^+) = \Theta(M^\al(J^{+-})) + \Theta(M^\al(J^{++})), \qquad J \in \tilde\I_N, \quad 0 \leq \al \leq m.$$
It remains to show that $\Theta^*$ preserves $d\mu_{A_\eps}$. For then if we do that, letting $\eps \to 0$ in (\ref{eq:WickExp2}) yields (\ref{eq:findimexp}).

Now $(\Theta(v), A_\eps \Theta(v)) = (v,A_\eps v)$ since the adjoint action preserves the inner product on the Lie algebra and our definition of $x^*$ ensures that elements of $\V_N$ which pair nontrivially always become conjugated by $h_{4N}$ evaluated at matching points. (In Figure 2, $x^*$ is the midpoint of $\bar x_i$ and $x_{i+1}$ for $J^+$, $J^{+-}$, and $J^{++}$.) It remains to show that $\Theta$ preserves the Lebesgue measure $\prod_{X \in \V_N} dX$, i.e., the Jacobian matrix $\J(\Theta)$ has determinant equal to one. We will show that $\J(\Theta)_{I, J}$ is block lower triangular (the blocks being given by the supressed curve indices and Lie-algebraic indices), with the diagonal blocks being unimodular. Here, we order our interval indices in ascending order as follows
$$J_1^-, J_1^+, J_1, J_2^-, J_2^+, J_2, \ldots, J_{2N}^-, J_{2N}^+, J_{2N}$$
where $J_i = [(i-1)/2N, i/2N] \in \I_{2N}$.

For $\Theta$ acting on $X=M^\alpha(J)$, $1 \leq \al \leq m$, then $\Theta(X)$ is linear in $X$. Hence the derivative of $\Theta(X)$ with respect to $X$ is just the adjoint action of $h_{4N}$, and this is unimodular on the $J$-block $\J(\theta)_{J,J}$ (the derivative of $\Theta(X)$ with respect to the $M^0$--variables comprising $h_{4N}$ will be in the strictly lower triangular part of $\J(\Theta)$). For the remaining case $X = M^0(J)$, $J \in \I_{4N}$, then $\Theta(X)$ is a formal power series in the $M^0(J')$, $J' \in \I_{4N}$, with $J' \leq J$ (in the natural ordering of intervals). Remarkably it is linear in $M^0(J)$, which yields that the corresponding diagonal block $\J(\Theta)_{J,J}$ is unimodular. This is evident for $J \in \tilde\I_N^{++}$, since then conjugation by $\ad(h_{4N,x^*})$ happens at the left-endpoint and so is comprised out of $M^0(J')$ with $J'$ strictly preceding $J$. For $J \in \tilde\I_N^{+-}$, let $J = [x,x+\Delta/4]$. So $x^* = x+\Delta x/4$, and we have
$$h_{4N,x^*} = (1 - M^0(J))h_{4N,x}.$$
Hence,
\begin{align*}
\Theta(M^0(J)) &= \ad(h_{4N,x^*})M^0(J)  \\
&= \ad(h_{4N,x})M^0(J)
\end{align*}
which is linear in $M^0(J)$. The lemma now follows.\End

\section{Proofs of Main Theorems}\label{Sec:Proofs}

We now prove our main theorems computing various kinds of expectations of the Wilson loop operator $W_{f,\gm}(A)$. As before, $\gm$ is an arbitrary piecewise $C^1$-closed curve which we may assume to be admissible, $\g$ is embedded inside $\mr{End}(V)$, the space of matrices acting on some vector space $V$, and $f = \tr$ (trace on $\mr{End}(V)$). Without loss of generality, we assume the image of $\gm$ is contained within the strip $[0,L] \times \R \subset \R^2$. Write
$$\gm \equiv \gm_m.\cdots.\gm_1$$
in terms of a horizontal curve decomposition. Let $x^\pm_\al$ be the initial and final $x$-coordinates of the horizontal $\gm_\al$, $1 \leq \al \leq m$. So these points satisfy the matching condition (\ref{eq:match}). Let $\gm_0: \R \to \R$ be the horizontal curve given by being the identity map along the $x$-axis in $\R^2$.

Let
\begin{align}
C^{\al\beta}(x) &= \sigma_{\al\beta}\bar G_{pax}(\bar \gm_\al(x),\bar\gm_\beta(x)) \label{eq:defC}
\end{align}
where $\sigma_{\al\beta} = \pm 1$ according to whether $\gm^\al$ and $\gm^\beta$ move in the same or opposite directions. Then
\begin{align}
\tilde C^{\al\beta}(x) &=  C^{\al\beta}(x) - C^{\al 0}(x) - C^{0\beta}(x) + C^{00}(x)\\
 &= \sigma_{\al\beta}\bar G_{ax}(\bar\gm^\al(x),\bar\gm^\beta(x)). \label{eq:tildeC2}
\end{align}
Thus, the ``random variables'' $M^\al(I)$ and $\tilde M^\al(I)$ as defined in Definition \ref{Def:M} and equation (\ref{def:tildeM}) from the previous section capture how the integrals of $A$ along $\gm^\al|_I$ are ``distributed'' in partial axial-gauge and complete axial-gauge, respectively, with only complete axial-gauge truly giving rise to honest measure theoretic notions.

Let
\begin{align}
 P^\al_N &= P^{M^\al}_{N, x^\al_- \to x^\al_+} \label{eq:PIto}\\
 \tilde P^\al_N &= \tilde P^{\tilde M^\al}_{N, x^\al_- \to x^\al_+} \label{eq:tildePStrat}
\end{align}
using Definition \ref{Def:PTIto} and \ref{Def:PTStrat}. They represent discretized It\^{o} and Stratonovich parallel transport along $\gm^\al$ in partial and complete axial-gauge, respectively. (Note that since $C^{\al\al}(x) \equiv 0$, we could have used Stratonovich parallel transport in (\ref{eq:PIto})).

We now prove Theorems \ref{Thm:1} and \ref{Thm:2}.

\begin{Theorem}
 Define
 \begin{equation}\left<W_{f,\gm}\right>_{ax} = \lim_{N \to \infty}\bE(\tr(\tilde P^{m}_N \cdots \tilde P^{1}_N)) \label{eq:defWax}
 \end{equation}
 Then
 $$\left<W_{f,\gm}\right>_{ax} = \left<W_{f,\gm}\right>$$
 for $\lam > 0$. Moreover, $\left<W_{f,\gm}\right>_{ax}$ defines an entire power series.
\end{Theorem}

\Proof In \cite{Dri}, it is shown how $\left<W_{f,\gm}\right>$, given by integrals involving heat kernels on $G$, can be computed by understanding the joint distribution of stochastic parallel transports of the constituent horizontal curves $\gm_\al$ of $\gm$. Namely, as a first step, if $\gm^\al$ is a right-moving horizontal curve, define $\tilde g_t^\al$ to be the solution to the  Stratonovich differential equation
\begin{align}
\begin{split}
 d\tilde g^\al_t + dW(\mathbf{1}_{R^\al(t)}) \circ \tilde g^\al_t &= 0 \\
 \tilde g^\al_{t_0} &= 1
 \end{split}\label{eq:StratSDE}
\end{align}
where (i) $t_0 = x_-^\al$; (ii) $W$ is two-dimensional $\g$-valued white-noise, i.e. $W = e_aW^a$ satisfies $\left<W^a(f_1), W^b(f_2)\right> = \lambda\delta^{ab}\left<f_1,f_2\right>_{L^2}$; (iii) $R^\al(t)$ is the region between the graph of $\bar\gm^\al|_{[x^\al_-, t]}$ and the $x$-axis.\footnote{This assumes the graph of $\bar\gm^\al$ lies above the $x$-axis. However, since partial axial-gauge and the lattice formulation are translation-invariant, we can suppose $\gm$ and hence the $\gm_\al$ all lie above the $x$-axis.} (For $\gm^\al$ left-moving, we define $\tilde g_t^\al$ by analogy:
\begin{align}
\begin{split}
 d\tilde g^\al_t + d_-W(\mathbf{1}_{R^\al(t)}) \circ \tilde g^\al_t &= 0 \\
 \tilde g^\al_{t_0} &= 1
 \end{split}\label{eq:StratSDE-left}
\end{align}
where $d_-$ is the backwards pointing differential and $R^\al(t)$ is defined as the region between the graph of $\bar\gm^\al|_{[x_+^\al,t]}$.) Note that (\ref{eq:StratSDE}) and (\ref{eq:StratSDE-left}) are true (continuous) parallel transport, not the time-discretized versions in the previous section. 

One of the main results of \cite{Dri} is that
\begin{equation}
 \left<W_{f,\gm}\right> = \bE\big(\tr(\tilde g^m_{x_+^m}\cdots \tilde g^1_{x_+^1})\big) \label{eq:Dri}
\end{equation}
where $\bE$ is an honest stochastic expectation. From the work of \cite{Ben}, the solution to (\ref{eq:StratSDE}) is given by the usual path-ordered exponential expansion, where all integrals are understood in terms of iterated Stratonovich integrals of the $dW(\mathbf{1}_{R^\al(t)})$. But from the definitions, we have $dW(\mathbf{1}_{R^\al(t)}) = \tilde M^\al_t$. So then the solution to (\ref{eq:StratSDE}) is given by
\begin{align}
\tilde g_t^\al &= 1 + \sum_{n=1}^\infty (-1)^n \int_{t \geq t_n \geq \cdots \geq t_1 \geq t_0}d\tilde M_{t_n} \circ \cdots \circ d\tilde M_{t_1}, \qquad 1 \leq \al \leq m  \label{eq:StratSDEexp} \\
&=: \sum_{n=0}^\infty [\tilde g_t^\al]_n,
\end{align}
where $[\tilde g_t^\al]_n$ denotes the $n$th summand of (\ref{eq:StratSDEexp}). 
Lemmas \ref{Lemma:Approx} and \ref{Lemma:StoI} show that
\begin{equation}
\bE\big(\tr(\tilde g^m_{x^m_+}\cdots \tilde g^1_{x^1_+})\big) = \lim_{N \to \infty}\bE\big(\tr(\tilde P^m_N \cdots \tilde P^1_N)\big) \label{eq:2E}
\end{equation}
since the $P^\al_N$ are discretizations of the $\tilde g^\al_N$. Thus, $\left<W_{f,\gm}\right> = \left<W_{f,\gm}\right>_{ax}.$

Finally, by substituting (\ref{eq:StratSDEexp}) into the right-hand side of (\ref{eq:Dri}), we show that (\ref{eq:Dri}) defines an entire power series in $\lambda$. This can be seen by estimating the coefficient of $\lambda^n$ arising from the term
\begin{equation}\sum_{n_m+\cdots +n_1=n}[\tilde g_{x^+_m}^m]_{n_m}\cdots [\tilde g_{x^+_1}^1]_{n_1},\label{eq:prod_iteratedS}
\end{equation}
and showing that it decays super-exponentially. Ultimately, this stems from the fact that support of an iterated integral of order $n$ decays as $1/n!$, which dominates an integrand that is exponentially bounded. More precisely, we proceed as follows:

First, any iterated Stratonovich integral of order $n$ (i.e. the $n$th order term in (\ref{eq:StratSDEexp})) can be re-expressed as a sum of $F_{n+1}$ many iterated Riemann-Ito integrals by repeated application of Lemma \ref{Lemma:StoI}, where $F_n$ is the $n$th Fibonacci number. (Here, we use the version of Lemma \ref{Lemma:StoI} obtained in the continuum limit $N \to \infty$, so that the last term of (\ref{eq:StoIto}) drops out.) This follows from the recurrence relation $F_{n+1} = F_n + F_{n-1}$, corresponding to the $(n+1)$-th iterated Stratonovich integral yielding two contributions: one coming from the $(n+1)$-th integral being contracted with the $n$th integral to become a single Riemann integral (there are $F_{n-1}$ many such terms) and one coming from no contractions taking place with the $(n+1)$th integral (there are $F_n$ many terms), which converts the latter to an Ito integral. The resulting $F_{n+1}$-many iterated Riemann-Ito integrals obtained in this manner is an iterated integral of order at least $n/2$.

Hence, each term $[\tilde g_{x^+_m}^m]_{n_m}\cdots [\tilde g_{x^+_1}^1]_{n_1}$ of the sum (\ref{eq:prod_iteratedS}) can be rewritten as a sum of $F_{n_m + 1}\cdots F_{n_1+1}$ many terms that are each a product of $m$ iterated Ito-Riemann integrals. Each such term is
of the schematic form
\begin{equation}
 \pm\left(\int_{x_m^+ > t_{n_m'} > \ldots > t_1 > x_m^-} X^m(t_{n_m'})\cdots X^m(t_1)\right)\cdots \left(\int_{x_1^+ > t_{n_1'} > \ldots > t_1 > x_1^-} X^1(t_{n_1'})\cdots X^1(t_1)\right) \label{eq:prodRI}
\end{equation}
where (i) the $X^\al$ variables are either $d\tilde M^\al$ or $\tilde C^{\al\al}$, $\al = 1,\ldots, m$; (ii) the order of the $\al$-th term of (\ref{eq:prodRI}) is $n_\al' := n_\al - k_\al$, where $0 \leq k_\al \leq n_\al/2$ is the number of $X^\al$'s equal to $\tilde C^{\al\al}$ (i.e. $k_\al$ is the number of contractions used in obtaining the $\al$-th term of (\ref{eq:prodRI}) from the iterated Stratonovich integral $[\tilde g^\al_{x_\al^+}]_{n_\al}$ ). The expectation of a term such as (\ref{eq:prodRI}) yields a product of iterated Riemann integrals obtained by Wick contracting all possible pair of $X$ variables that are of type $d\tilde M^\al$, i.e. replacing $X^\al(t)$ and $X^\beta(t)$, $\al \neq \beta$ at equal times with $\tilde C^{\al\beta}(t)$. 

The resulting product of integrals, call it $\mc{R}$, can be bounded in terms of three factors: (i) the volume of the supports of each of the $m$-many integrals occurring in (\ref{eq:prodRI}) ; (ii) a pointwise bound on $\tilde C^{\al\beta}$; (iii) a combinatorial factor for all the possible Wick contractions. For (i), if we let $L \geq 1$ be large enough so that the support of $\gm$ is contained in a box of horizontal length $L$, then the product of volumes we obtain is
$$\prod_{\al=1}^m \frac{L^{n_\al'}}{n_\al'!} \leq \frac{L^{n}}{(n/2)!}$$
since $n_\al' \geq n_\al/2$. For (ii), if we let $C = \max(1,\sup_{\al,\beta, t} C^{\al\beta}(t))$, then the $n/2$-many factors of $C^{\al\beta}$ occurring in $\mc{R}$ can be bounded by $C^{n/2}.$ Finally for (iii), the combinatorial factor we obtain is (very crudely) bounded by $(m-1)!!^{n/2}$, since from the $m$-fold product (\ref{eq:prodRI}) at most $m$ many $X$'s can have coinciding times (they must occur from distinct factors) and there are no more than $n/2$ possible Wick contractions. Here $(m-1)!!$ arises from the formula
$$\frac{1}{\sqrt{2\pi}}\int_\R e^{-x^2/2}x^m dx = \begin{cases}
(m-1)!! & m \textrm{ is even}\\                                                  
0       & m \textrm{ is odd}                                          \end{cases}$$
yielding the number of ways to completely Wick contract a set of $m$ slots. Hence, we see that the expectation of (\ref{eq:prodRI}) is bounded by $\frac{K^n}{(n/2)!}$ where $K$ is a sufficiently large constant independent of $n$. Since the term 
$[\tilde g_{x^+_m}^m]_{n_m}\cdots [\tilde g_{x^+_1}^1]_{n_1}$ can be expressed as exponentially-many (in $n$) terms of the form (\ref{eq:prodRI}), and there are at most $n^m$ solutions to $n_1 + \ldots + n_m = n$, it follows that we can bound the expectation of the $\lambda^n$ term of (\ref{eq:prod_iteratedS}) by $\frac{K^n}{(n/2)!}$ for $K$ sufficiently large. It follows that $\left<W_{f,\gm}\right>_{ax}$, which equals the sum over all $n$ of the expectations of (\ref{eq:prod_iteratedS}), defines an entire power series in $\lambda$.\End

Note that the left-hand side of (\ref{eq:2E}) is a priori a nonexplicit stochastic expectation. On the other hand, the right-hand side of (\ref{eq:2E}), for each $N$, can be evaluated using the Wick rule. Letting $N \to \infty$, the expectation converges to a (complicated) sum of Riemann integrals yielding $\left<W_{f,\gm}\right>_{ax}$. See Remark \ref{Rem:ax} below for an example computation.\\

\noindent \textbf{Proof of Theorem \ref{Thm:2}}: We have
\begin{equation}
\left<W_{f,\gm}\right>_{pax} = \lim_{N \to \infty}\bE\big(\tr(P^m_N \cdots P^1_N)\big),
\end{equation}
which one can either take as a definition or else see that it is equivalent to the definition we gave in the introduction involving Feynman diagrams using the partial axial-gauge propagator. Next, we can replace each $P^\al_N$, which is the element $g^\al_{N,x_+^\al}$ given by (\ref{eq:defg}), with $\hat g^\al_{N,x_+^\al}$ as defined in (\ref{eq:hatg}), due to conjugation-invariance of $\tr$. We then replace $\hat g^\al_{N,x_+^\al}$ with the leading term of (\ref{eq:tildegdecomp}) by Lemma \ref{Lemma:ChangeGauge}, since the terms of order $o(1)$, as defined by Definition \ref{Def:order}, vanish in the limit $N \to \infty$. Next, we apply Lemma \ref{Lemma:Changevar} to eliminate all the $\ad(h_{4N})$ terms in the leading terms of (\ref{eq:tildegdecomp}) when we compute the expectation. But once we remove the $\ad(h_{4N})$ terms, what we have left is Stratonovich parallel transport with respect to the $\tilde M^\al$, i.e., we obtain
$$\left<W_{f,\gm}\right>_{pax} = \lim_{N \to \infty}\bE\big(\tr(\tilde P^m_N \cdots \tilde P^1_N)\big).$$
This shows $\left<W_{f,\gm}\right>_{pax} = \left<W_{f,\gm}\right>_{ax}$.\End

\begin{Remark}\label{Rem:Massless}
 Suppose we insert a mass $m^2$ into the Yang-Mills action in partial axial-gauge
 \begin{equation}
  \frac{1}{2\lambda}\int dxdy\Big(\left<\pd_y A_x,\pd_y A_x\right> + m^2 \left<A_x,A_x\right>\Big).
 \end{equation}
Replacing the action in (\ref{eq:YMpax}) with the one above, we obtain a bona fide Gaussian measure whose covariance is $\lambda$ times
\begin{equation}
G_m(x-x',y-y') = \frac{e^{-m|y-y'|}}{2m}\delta(x-x'). \label{eq:mG}
\end{equation}
Note that $\frac{1}{2m}e^{-m|y-y'|}$ is the unique  Green's function for $-\pd_y^2+m^2$ that is invariant under translations and reflections. Thus, we obtain a measure which is Ornstein-Uhlenbeck measure in the $y$-direction and white-noise in the $x$-direction. Unfortunately, the $m^2 \to 0$ limit of (\ref{eq:mG}) does not exist. (This is because the limit kinetic operator has constant zero modes, which if taken into account, would yield a sensible limit via renormalization by an additive constant:
\begin{equation}
\lim_{m \to 0}\frac{e^{-m|y-y'|} - 1}{2m} = -\frac{1}{2}|y-y'|. \label{eq:subtract}
\end{equation}
This limit recovers the partial axial-gauge Green's operator.)

Nevertheless, we can try to analyze stochastic holonomy with respect to (\ref{eq:mG}) and let $m \to 0$. Unfortunately, for nonabelian gauge group $G$, it is unclear whether the limiting (expectation of) holonomy exists. For $G$ abelian, we were able to organize Feynman diagrams in a way that exhibits the zero mode subtraction (\ref{eq:subtract}) so that the $m \to 0$ exists. Unfortunately, when $G$ is nonabelian, the noncommutativity of the basis elements $e_a$ of $\g$ complicates the combinatorics involved in showing that singular elements cancel in the massless limit, see the next remark. Being unable to control such combinatorics for general curves to arbitrary order, we instead developed the algebraic stochastic methods in this paper. In some sense, our algebraic use of gauge-invariance in Lemma \ref{Lemma:Changevar}  circumvents such difficult combinatorics. See also Remark \ref{Rem:c} for another attempt to restore an honest measure-theoretic setting.
\end{Remark}

\begin{Remark}\label{Rem:ax}
 The subtraction (\ref{eq:subtract}) also suggests how Feynman diagrams should be organized when equating $\left<W_{f,\gm}\right>_{ax}$ with $\left<W_{f,\gm}\right>_{pax}$. Let us sketch how to compute $\left<W_{f,\gm}\right>_{ax}$ for the example in the introduction. We proceed somewhat formally using the direct expression (\ref{eq:Pax}) to compute integrals; these statements can be justified by using the rigorous stochastic definition (\ref{eq:defWax}).

 We have two types of Wick contractions for $\left<W_{f,\gm}\right>_{ax}$. As before, we have those which are vertical ``chords'' joining a pair points on $\gm_1$ and $\gm_3$. But now we also have ``tadpoles'' joining adjacent pairs of points on either $\gm_1$ or $\gm_3$ (which makes their $x$-coordinates collapse). (If we Wick contract nonadjacent points of $\gm_1$ and $\gm_3$, the path ordering condition makes intermediate points range over a set of measure zero, and thus we obtain zero.) Tadpoles contribute a factor of $+\frac{1}{2}\bar G_{ax}(y,y) = \frac{|y|}{2}$; the plus sign arises because we are joining two points along the same curve, and the $\frac{1}{2}$ arises from a careful analysis of the Stratonovich midpoint rule. Thus, ignoring Lie-algebraic factors for the moment, we can organize Wick contractions into triplets consisting of a tadpole on a double point $t_i$ belonging to $\gm_1$, a chord joining $t_i$ to $t_{i+n}$, and a tadpole on a double point $t_{i + n}$ belonging to $\gm_3$. (Thus, in Figure 1, we decorate each chord with a tadpole and obtain a ``dumbbell''.) The scalar part of each such dumbbell yields
 $$-\bar G_{ax}(\gm_+(x_i), \gm_-(x_i)) + \frac{1}{2}\bar G_{ax}(\gm_-(x_i), \gm_-(x_i)) + \frac{1}{2}\bar G_{ax}(\gm_+(x_i), \gm_-(x_i)) = \frac{1}{2}(\gm_+(x_i) - \gm_-(x_i)),$$
 exactly coinciding with the scalar contribution of the partial axial-gauge propagator. Fortunately, for the simple convex curve we have drawn, the Lie-algebraic factors all become powers of the quadratic Casimir, and so all  scalar factors are weighted equally. Thus, the above computation implies that $\left<W_{f,\gm}\right>_{ax} = \left<W_{f,\gm}\right>_{pax}$ at every order in $\lambda$.

 However, as soon as $\gm$ begins to wind nontrivially or self-intersect, we obtain nontrivial Lie-algebraic factors that weight various Wick contractions. It becomes nontrivial to organize these factors in a way that exhibits the equality between $\left<W_{f,\gm}\right>_{ax}$ and $\left<W_{f,\gm}\right>_{pax}$. If we had used the mass regulated propagator (\ref{eq:mG}), such Lie-theoretic factors are the source of the combinatorial difficulty involved in showing that a massless limit exists.
\end{Remark}

\begin{Remark}\label{Rem:c}
 Given a constant $c > 0$, consider the following Green's function
 \begin{equation}
\bar G_{pax + \frac{1}{2}c}(y,y') = -\frac{1}{2}|y-y'| + \frac{1}{2}c. \label{eq:Gpaxc}
 \end{equation}
Given any bounded domain $D \subset \R$, for sufficiently large $c$, $\bar G_{pax + \frac{1}{2}c}(y,y')$ will define a positive
integral-kernel pairing for functions supported on $D$. (In fact, we can take $D = [-c,c]$). It follows that if we define Wilson loop expectations $\left<W_{f,\gm}\right>_{pax + \frac{1}{2}c}$ using $\bar G_{pax + \frac{1}{2}c}$ in place of $\bar G_{pax}$, with $c$ sufficiently large so that the support of $\gm$ is contained in the strip $\R \times [-c,c]$, we place the expectation $\left<W_{f,\gm}\right>_{pax + \frac{1}{2}c}$ back into an honest measure-theoretic setting. Moreover, if instead of (\ref{eq:defC}), we make the definition
$$C^{\al\beta}(x) = \sigma^{\al\beta}\bar G_{pax+\frac{1}{2}c}(\bar\gm^\al(x),\bar\gm^\beta(x)),$$
then
$$\tilde C^{\al\beta}(x) = \sigma^{\al\beta}\bar G_{ax}(\bar\gm^\al(x),\bar\gm^\beta(x)),$$
agreeing with (\ref{eq:tildeC2}). Consequently, repeating the steps of the proof of Theorem 2 shows that
\begin{equation}
\left<W_{f,\gm}\right>_{pax + \frac{1}{2}c} = \left<W_{f,\gm}\right>_{ax}. \label{eq:paxc=ax}
\end{equation}

While this procedure avoids the algebraic stochastic analytic setup which is the main development of this paper\footnote{The author thanks an anonymous referee for raising this point.}, this circumvention is unsatisfactory for a number of reasons. First of all, it only establishes (\ref{eq:paxc=ax}) for sufficiently large $c$ and does not address the case $c=0$, which is the most natural choice and the one used in the existing literature when employing the Feynman diagram method (e.g. \cite{BasNar96, KK}). Second, it is unnatural to impose a choice of $c$ at the outset, since it puts a restriction on the kind of observables one is willing to consider. For instance, it is of interest to consider Wilson loop asymptotics as the underlying loop $\gm$ tends to infinity; this kind of analysis is limited if we impose $\gm \subset \R \times [-c,c]$. Finally, the aforementioned procedure is at odds with our main motivation of analyzing quantum-field theoretic quantities that lie outside a proper measure-theoretic framework.

\end{Remark}

\end{document}